\documentclass[a4paper,fleqn,usenatbib]{mnras}
\usepackage{cleveref}
\usepackage{newtxtext,newtxmath}
\usepackage[T1]{fontenc}
\usepackage{ae,aecompl}
\usepackage{graphicx,amsmath,amssymb}
\usepackage{booktabs}
\usepackage{todonotes}

\newcommand{\EE}{\mathbf{E}}
\newcommand{\BB}{\mathbf{B}}
\newcommand{\II}{\mathrm{i}\,}
\renewcommand{\Re}{\mathrm{Re}\,}
\renewcommand{\Im}{\mathrm{Im}\,}
\newcommand{\res}{\mathrm{res}}
\renewcommand{\vec}{\mathbf}

\crefname{figure}{Fig.}{Figs.}
\Crefname{figure}{Fig.}{Figs.}
\crefformat{appendix}{#2Appendix#3}
\Crefformat{appendix}{#2Appendix#3}

\title[Saturation of parallel ion instabilities]{How it cramps the flow: three regimes for the saturation of parallel ion-beam instabilities}
\author[M. S. Weidl et al.]{%
    M.~S.~Weidl,$^1$\thanks{E-mail: mwei@physics.ucla.edu (MSW)}
    D.~Winske$^2$,
    and C.~Niemann$^1$
    \\
    $^1$ UCLA Department for Physics \& Astronomy, 475 Portola Plaza, Los Angeles, CA 90095, USA
    \\
    $^2$ Los Alamos National Laboratory, Los Alamos, NM 87545, USA
}
\date{\today}
\pubyear{2016}

\begin{document}
\label{firstpage}
\pagerange{\pageref{firstpage}--\pageref{lastpage}}
\maketitle

\begin{abstract}
Motivated by recent advances in laboratory experiments on parallel ion-beam instabilities, we present a theoretical framework for and simulations of their evolution towards shock formation and Fermi acceleration. After reviewing the theory of beam instabilities with a focus on the so-called nonresonant or Bell instability, which we show to be due to the gyromotion of background ions, we contrast the saturation of three parameter regimes: (I) the left-handed `nonresonant' regime, (II) the right-handed beam-gyroresonant regime, (III) the balanced, mixed-turbulence regime.

\end{abstract}

\begin{flushright}
\emph{I will never write another letter with alternate inks. You cannot imagine how it cramps the flow of the style.}\\-- Charles Lamb to William Wordsworth, June 7, 1819
\end{flushright}

\section{Introduction}

From the solar wind to the jets of active galactic nuclei, astrophysical plasmas of all scales are constantly being stirred by counterstreaming flows. A large portion of the theory of plasma physics is therefore dedicated to the waves excited in such systems, and their marginal stability in weakly damped plasmas has been comprehensively understood for several decades \citep{krall1973,akhiezer1975}. However, the saturation of these modes is inextricably connected to, and arguably synonymous with, the highly nonlinear subject of plasma turbulence and still stimulates active debate in the astrophysical community, e.g.\ the saturation of the so-called nonresonant cosmic-ray instability \citep{bell04}.

Early in the history of solar-wind research, space physicists recognised how important electromagnetic waves created by ion beams could be for parallel collisionless shocks \citep{cipolla77,sentman81}. Countless satellite measurements from upstream of planetary bow shocks, where ions are reflected and accelerated by magnetic turbulence, have been published and analysed \citep[e.g.][]{fairfield69,russell79,lucek08,jian14}. Motivated by these observations and space experiments with artificial comets \citep{gleaves88,sauer99}, theoretical and numerical studies of parallel shocks soon began to focus on two electromagnetic instabilities, termed the right-hand resonant instability (RHI) and the nonresonant instability (NRI). \citet{gary91} extensively reviews these and a host of other modes from a space-physics perspective. In recent years the formation of collisionless shocks has also been studied in laboratory experiments \citep{huntington15, schaeffer17}.

Meanwhile, in order to explain cosmic-ray (CR) energies of the knee energy $E\sim10^{15}$ eV, astrophysicists considered collisionless shocks with much higher Alfv\'enic Mach numbers ($M_A\sim10^3$ as opposed to $M_A\lesssim10$ for the terrestrial bow shock) as the sites of diffusive shock acceleration or DSA \citep{axford77,bell78,blandford78}. Since the most efficient environment for DSA is a parallel collisionless shock \citep{caprioli14}, the magnetic turbulence that accelerates knee-energy CRs is likely generated by parallel beam instabilities coupling the most energetic debris of supernova remnants (SNR) with the surrounding interstellar medium. Commonly studied in this context are the gyroresonant CR streaming instability \citep{kulsrud69} and the so-called nonresonant CR instability \citep{bell04}.

The latter mode has been investigated extensively in recent years since it grows more rapidly than the streaming instability discovered by Kulsrud, at least for fast CR particles and a cold interstellar medium. Although it is sometimes referred to as a magnetohydrodynamic (MHD) instability, all of its analytical derivations depart from the kinetic Vlasov equation for a parallel-beam configuration \citep{achterberg83,bell04,amato09,zweibel10}. Early MHD simulations added a static CR current and a charge-compensating return current to Amp\`ere's law and showed significant magnetic-field growth, which saturated at values of $\delta B/B_0 \sim 10$ because of fieldline tension \citep{bell04,zirakashvili08}. \citet{pelletier06} argued that the instability would saturate because of a nonlinear kinetic effect, i.e.\ fieldline filaments growing to diameters comparable to the CR gyroradius. Computational astrophysicists who ran fully kinetic particle-in-cell (PIC) simulations had to choose between resolving the small-wavelength Bell mode or containing all growing long-wavelength modes in the simulation domain and understandably opted for the former. Starting with a monoenergetic beam of CR quasiparticles, these simulations showed that field growth saturated at $\delta B/B_0\lesssim10$ because of a rapid decrease in the relative drift velocity: as the cosmic-ray particles decelerated, the background ions quickly accelerated \citep{niemiec08,riquelme09}. Simultaneously the dominant wavelength was growing, seemingly without bound, and the ions began to form density filaments. A deceleration of cosmic-rays was also observed by \citet{lucek00}, using a hybrid code that treated the background plasma as a MHD fluid and the cosmic-ray particles kinetically.

A different kind of hybrid code has been a workhorse of the space-physics community since the early 1980s: modeling both background and beam ions by the PIC method and electrons as an inertialess MHD fluid, \citet{winske84} first investigated the NRI in numerical simulations, but with a focus on long-wavelength modes. \citet{quest88} used the same simulation technique to contrast the role of the left-handed NRI, which he recognised as being closely related to the kinetic firehose instability, with the role of the RHI in the formation of parallel collisionless shocks. After pointing out the existence of a short-wavelength electron-ion-whistler instability \citep{akimoto87}, \citet{akimoto93} ran hybrid simulations of the saturation of the RHI and NRI at $\delta B/B_0\sim4$ with $M_A=10$. Computational constraints limited all of these early simulations to resolving the one dimension along the initial magnetic field.

More recently, 2D hybrid simulations with much higher Mach numbers ($M_A>10^4$ in \citet{gargate10}), based on the same kinetic-ion/MHD-electron model, confirmed that Bell's instability saturates at $\delta B/B_0\sim10$ due to a decreasing relative beam drift if the feedback of turbulence on the CR beam is included. Later hybrid simulations, starting with a cold CR beam, showed that a right-handed mode dominates at lower Mach numbers and that the left-handed Bell instability only grows faster for high CR currents \citep{gargate12}, and that MFA could even reach $\delta B/B_0\sim10^2$ locally within the forming density filaments \citep{caprioli13}, but never explained the physical mechanism that simultaneously decelerated the CRs and accelerated the background ions. This coupling, which ultimately causes the saturation of Bell's instability, is the subject of this article.

We first sketch the analytical derivation of parallel ion-beam instabilities in \cref{secTheory} and compare exact solutions of the linear dispersion relation to commonly used approximations. \Cref{secGrowth} introduces the left-handed regime, the right-handed regime, and the mixed regime and compares the early stage of 2D hybrid simulations for each regime with analytical theory. In \cref{secSaturation}, we contrast the mechanisms leading to saturation in the nonlinear phase of each of the three regimes and how the ion populations are affected. Astrophysical implications for SNR shock acceleration are discussed in \cref{secDiscussion}. A detailed algebraic derivation of the `nonresonant' instability and its gyroresonant character is included in the \cref{appendix}.

\section{Linear theory}
\label{secTheory}

\subsection{Signed frequencies and other conventions}

\begin{figure}
\centering
\includegraphics[scale=.95]{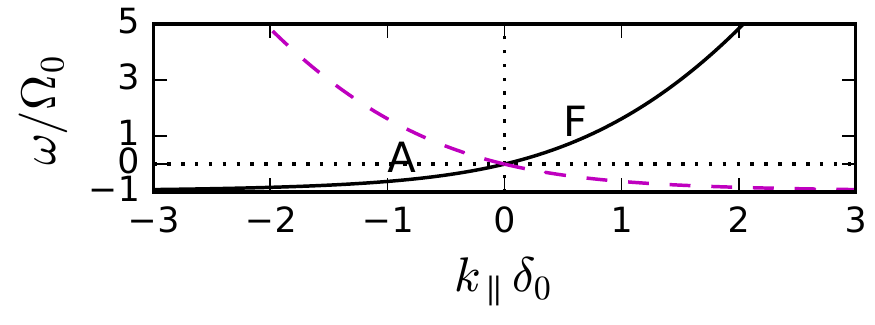}
\caption{Dispersion relation~\eqref{eqnDR} for left- and right-handed Alfv\'en waves propagating to the right (solid, $\omega/k_\parallel>0$) and to the left (dashed, $\omega/k_\parallel<0$). Frequency regimes of the rightwards fast magnetosonic (F) and Alfv\'en (A) wave are indicated.}
\label{figAlfvenDisp}
\end{figure}

Before we introduce various moving frames, it is crucial to understand what the signs of frequency and wavenumber imply in a single frame of reference. Hence this subsection summarises the sign conventions, following e.g.\ \citet{gary91}.

We first consider an electron-ion plasma with a unidirectional magnetic field $\BB_0 = B_0\, \mathbf{\hat z}$, such that $\beta = v_{\mathrm{th}}^2/v_A^2 \ll 1$. Here the thermal and Alfv\'en velocities are denoted as $v_{\mathrm{th}}$ and $v_A$, respectively. As we are only interested in the dynamics of ions in this article, we will make the standard hybrid-model assumption of strongly magnetised, inertialess electrons ($m_e\to0$, $|m_e\Omega_e|\to e\,B_0$) that compensate any charge imbalance far more quickly and gyrate much faster than any ion could react.

In this configuration, the plasma can become unstable to electromagnetic (EM) waves that propagate parallel to $\BB_0$ with frequencies $\omega$ close to the background-ion gyrofrequency, $-\Omega_0 \lesssim \omega \ll |\Omega_e|$. We designate left-handed polarisation around the magnetic-field direction, i.e.\ the polarisation of the ion gyromotion, with a negative frequency. This corresponds to the standard Fourier sign convention with $\exp(+\II k_\parallel z - \II \omega t)$ and expresses in which direction the perpendicular magnetic-field vector rotates if one measures it at a fixed position in space --- `fixed' in a frame of reference that must be specified. If all ions and electrons share one common rest frame, that frame is the obvious choice. Thus we distinguish fast magnetosonic/whistler waves (with right-handed circular polarisation and positive frequency) and the shear-Alfv\'en/ion-cyclotron waves (with left-handed circular polarisation and negative frequency) by the sign of $\omega$. The solid lines in \cref{figAlfvenDisp} show how the dispersion branches for both forwards-propagating modes merge into one continuous branch with this convention, which is described by the relation \citep[e.g.][]{kulsrud2005}
\begin{equation}
k_\parallel^2 = \frac{\omega^2}{c^2}\left(1+\frac{\omega_p^2}{\Omega_0\,(\omega+\Omega_0)}\right)
\label{eqnDR}
\end{equation}
and the condition $\omega/k>0$ for rightwards propagation.

In contrast to polarisation, helicity describes the direction in which a magnetic fieldline winds around the $z$ axis if one traces it in space at a fixed point in time. Helicity is determined from the sign of $k_\parallel$ that dominates in its Fourier spectrum: modes with positive helicity result in a right-handed fieldline as one follows its trajectory at a fixed point in time, while $k_\parallel<0$ corresponds to left-handed helicity. Whereas helicity is invariant under proper Lorentz transformations (which can change the norm but not the sign of $k_\parallel$) and thus well-suited for characterising modes independently of the reference frame, the polarisation depends on the frame one decides or is forced to work in. Resonant coupling of a wave to the gyromotion of an ion species (as for an ion-cyclotron wave) occurs only if $\omega' \approx -\Omega_g$, where $\omega'$ is the wave frequency Doppler-shifted to the rest frame of the ion with gyrofrequency $\Omega_g$:
\begin{equation}
\omega' = \omega - k_\parallel\, v_\parallel \approx - \Omega_g,
\label{eqnGyroresonance}
\end{equation}
where $v_\parallel$ is the parallel velocity of the ion in the same frame of reference in which $\omega$ is the measured wave frequency.

Although helicity (i.e.\ $k_\parallel$) is useful to describe a mode independently of the reference frame, the frame-dependent polarisation (i.e.\ $\omega'$) is not. Speaking of polarisation as a property of a mode or using it to determine gyroresonance is only useful if the reference frame is clear from the context.

\subsection{Cold interacting beams}

Let us now assume that a second species of beam ions with gyrofrequency $\Omega_b$ streams along the $z$ axis with velocity $\vec V_b$ relative to the background ions, or velocity $v_b$ in an arbitrary frame of reference. Writing $v_0$ for the parallel velocity of the background ions and $v_e$ for the electrons, the cold-plasma dispersion relation (derived in the \cref{appendix}) becomes in this arbitrary frame
\begin{multline}
	\left( \omega^2 - k_\parallel^2~c^2 \right)~\frac{n_e}{\omega_p^2} =\\
	n_0~\frac{\omega - v_0~k_\parallel}{\omega - v_0~k_\parallel + \Omega_0} - n_e~\frac{\omega - v_e~k_\parallel}{\Omega_0} + n_b~\frac{q_b^2}{m_b}~\frac{\omega - v_b~k_\parallel}{\omega-v_b~k_\parallel+\Omega_b},
\label{eqnBeamDR}
\end{multline}
where $\omega_p = n_e\,e^2/m_0$ is the reference plasma frequency, $n_e$ is the electron density, and $q_b$ and $m_b$ are beam-ion charge and mass, respectively normalised to the background-ion charge $q_0=e$ and mass $m_0$. The Alfv\'en velocity is defined with respect to the total electron density $n_e$, such that $v_A/c=\Omega_0/\omega_p$; the ion inertial length is $\delta=c/\omega_p$. The background- and beam-ion velocities, $\vec v_0$ and $\vec v_b = \vec V_b + \vec v_0$, as well as the drift velocity $\vec v_e = (n_0\,\vec v_0+q_b\,n_b\,\vec v_b)/n_e$ of the inertia-less electron fluid, are all parallel to the external magnetic field. Thus we assume that the drift of the electron population has established a return current that compensates the electric current due to the beam ions. Under the hybrid-model assumptions, the temperature of the electron population can be shown to be irrelevant for the growth of parallel propagating modes.

\begin{figure}
\centering
\includegraphics[scale=.95]{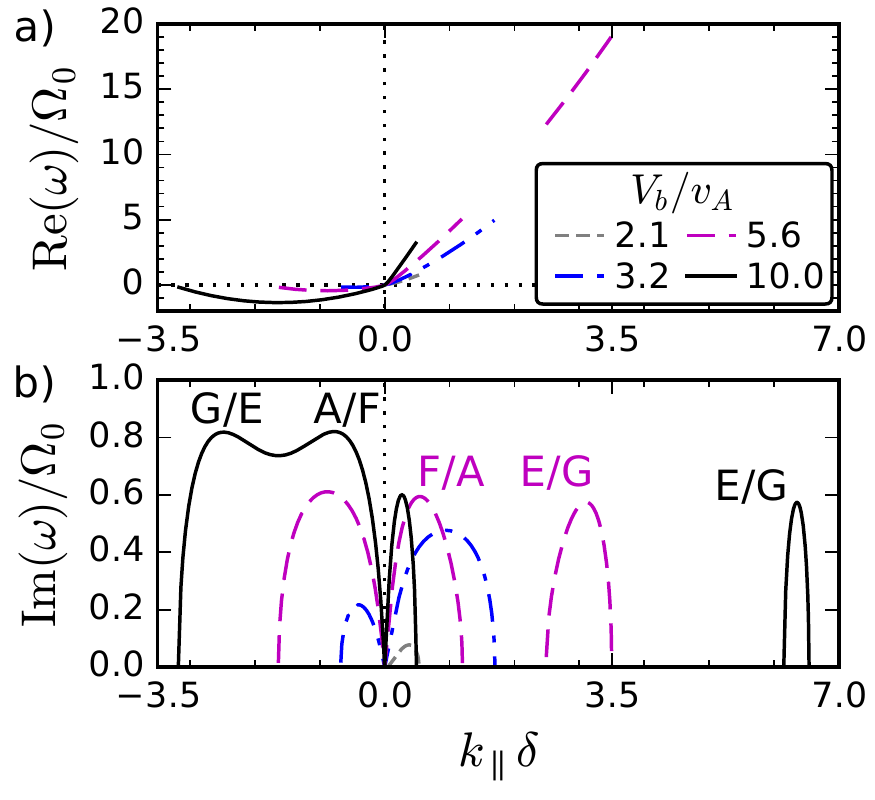}
\caption{a) Unstable frequency (in rest frame of background ions) and b) growth rate for a cold ion beam in a background plasma ($n_b/n_e = 0.33$) with increasing relative drift velocity $V_b$. The E/G mode for $V_b=10\,v_A$ lies at $\Re(\omega)\gtrsim60\,\Omega_0$ and is not shown in the top plot.}
\label{figVlasovLR}
\end{figure}

Since we are interested in parallel modes that are absolutely growing, not spatially aperiodic convective instabilities, we limit the domain of solutions to $k \in \mathbb{R}$ and $\omega \in \mathbb{C}$ with $\Im\omega\geq0$. The dispersion relation for two cold ion beams becomes a simple fourth-degree algebraic curve in $\omega$ and $k_\parallel$ after multiplying both sides by the denominators of the right-hand side and can thus be solved exactly with a computer algebra system. Exact solutions for the real frequency in the background-ion frame and for the growth rate are shown in \cref{figVlasovLR} for $c=2000\,v_A$.

As one increases $V_b$, the unstable modes, or local maxima of $\Im\omega$ in \cref{figVlasovLR}b, become more numerous: starting with no unstable modes at $V_b<2\,v_A$, the system becomes weakly unstable ($\Im\omega<0.1\,\Omega_0$) to a single mode with positive helicity $0<k_\parallel<0.5\,\delta$ and frequency at $V_b=2.1\,v_A$. A second mode, this time with negative helicity and frequency, exists for $V_b=3.2\,v_A$ (blue dash-dotted). Increasing the drift further to $V_b=5.6\,v_A$ (magenta dashed), the right-handed mode splits into two, one at low $k_\parallel$ (marked with F/A) and one with $k_\parallel\delta>1$ (E/G), in addition to the single left-handed mode. The latter begins to split at $V_b=10\,v_A$ (black solid), revealing a left-handed high-wavenumber mode (G/E) and a left-handed low-wavenumber mode (A/F).

Hence the system can be unstable to up to four distinct modes, two with positive helicity, two with negative helicity. The polarisation of these modes matches their helicity in the background frame, as all components have momentum in the positive $x$ direction, or in the case of background ions no momentum at all, and can therefore only excite waves with $\omega/k_\parallel>0$. Note that all negative-helicity modes in this frame have $-2\,\Omega_0<\omega<0$ and thus fulfill the gyroresonance condition~\eqref{eqnGyroresonance} to within one gyrofrequency, i.e.\ they are gyroresonant with the background ions. We again refer to the \cref{appendix} for analytical arguments. Doppler-shifted to the beam frame, the polarisation of all waves is reversed and $\omega/k_\parallel<0$ for all modes. Plotting the Doppler-shifted frequencies $\Re(\omega')$ of the modes with $k_\parallel>0$, we would find $-2\,\Omega_0<\Re(\omega')<0$ in the beam frame. In the rest frame of the (inertialess) electrons, all frequencies are positive, corresponding to the right-handed polarisation of the electron gyromotion.

In an attempt to introduce some systematic, unifying terminology, we will call long-wavelength modes with $|k_\parallel\delta|<1$ fast magnetosonic (F) if their polarisation in a certain frame is right-handed and shear-Alfv\'enic (A) if it is left-handed (cf.\ \cref{figAlfvenDisp}). Similarly, short-wavelength modes are attributed to electron-whistler waves (E) if right-handed and to the ion gyromotion (G) if left-handed.

\begin{figure}
\centering
\includegraphics[scale=.95]{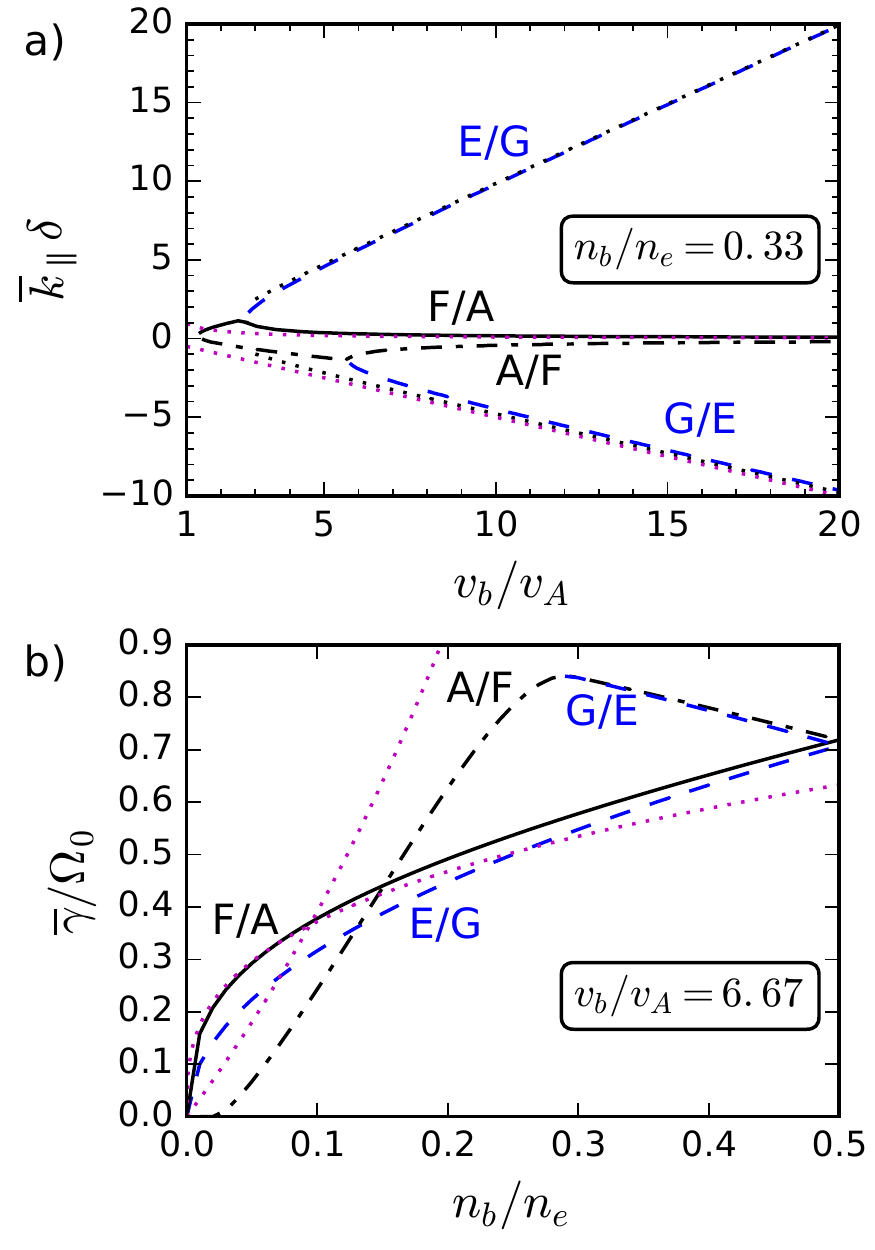}
\caption{a) Wavenumbers $\overline k_\parallel(v_b)$ of fastest-growing modes as predicted by the cold-plasma dispersion relation~\eqref{eqnBeamDR} with $n_b/n_e=0.33$ and $q_b=m_b=1$. b) Growth rates $\overline \gamma(n_b)$ of fastest-growing modes for a fixed beam velocity $v_b=6.67~V_A$ in the electron frame. Dotted lines represent the analytical predictions according to equations~\eqref{eqnGaryRHI} and \eqref{eqnWinskeNRI}.}
\label{figKmax}
\end{figure}

Descending from high to low $k_\parallel$, we then interpret the four modes as a right-handed electron-whistler wave driven by the beam-ion gyromotion (E/G), a fast magnetosonic mode in the background plasma in resonance with a shear-Alfv\'en wave in the beam plasma (F/A), a shear-Alfv\'en mode in the background resonating with a fast magnetosonic mode in the beam (A/F), and an ion-cyclotron wave in the background, driven by the gyration of the background ions, which becomes an electron-whistler wave in the beam frame (G/E). \Cref{figKmax}a shows for which $k_\parallel$ the growth rate attains a local maximum, and thus how the E/G and F/A modes separate at high enough velocities, and how G/E splits off from A/F at even higher values of $v_b$.

The motivation for this --- at first glance admittedly somewhat cryptic --- nomenclature is to emphasise the frame-dependence of polarisation and the underlying symmetry of the modes. A mode that appears to be a left-hand polarised long-wavelength mode (i.e.\ an Alfv\'en wave A) in the rest frame of the background ions turns into a right-hand polarised long-wavelength mode (i.e.\ a fast magnetosonic wave F) in the rest frame of the beam ions. When we later investigate how background and beam ions are affected by waves, it will be important to keep both perspectives in mind at all times. The above mode is therefore termed an A/F mode, as it primarily accelerates the background like an A wave and scatters the beam like a F wave. Similarly, an E/G mode looks like a short-wavelength right-handed electron whistler (E) in the background frame, but becomes in the beam frame a short-wavelength mode that is resonant with the ion gyromotion (G). Readers that prefer to focus on the background frame can simply ignore the second, `beam-frame' letter in our mode terminology.

In the space-physics community, the resonant right-hand instability or RHI \citep{sentman81} is a well-known mode with positive helicity. Using a number of simplifications (viz.~$\Re\omega_{\mathrm{RHI}} = k_{\mathrm{RHI}}\, v_b - \Omega_b$, $|\omega_{\mathrm{RHI}}|\ll\Omega_0=\Omega_b$, $v_B\gg V_A$, and $q_b\,n_b\ll n_e$) and calculating in the electron frame, \citet{gary78} found for the growth rate $\gamma_{\mathrm{RHI}} = \Im\omega_{\mathrm{RHI}}$ and the wavenumber $k_{\mathrm{RHI}}$ of the fastest-growing RHI mode
\begin{equation}
	\overline \gamma_{\mathrm{RHI}} \sim \left( \frac{q_b\,n_b}{2\,n_e} \right)^{1/3}~\Omega_0,\quad \overline k_{\mathrm{RHI}} \sim \left[1-\left(\frac{q_b\,n_b}{2\,n_e}\right)^{1/3}\right]~\frac{\Omega_0}{v_b}.
	\label{eqnGaryRHI}
\end{equation}
As $0<\overline k_{\mathrm{RHI}}\,\delta<1$, this instability is identical to our F/A mode.

With a similar set of assumptions, but additionally demanding $q_b\, n_b\, v_b\leq n_e\, v_A$ and $q_b\, n_b\, v_b^2\gg n_e\, v_A^2$, \citet{winske84} found for the growth rate of a left-handed mode called the nonresonant instability or NRI:
\begin{equation}
	\overline \gamma_{\mathrm{NRI}} \sim \frac{q_b\,n_b/n_e}{2\,\left(n_0/n_e\right)^{3/2}}~\frac{v_b}{v_A}~\Omega_0,\quad \overline k_{\mathrm{NRI}} \sim - \frac{q_b\,v_b}{2\,v_A} \frac{\Omega_0}{v_A}.
	\label{eqnWinskeNRI}
\end{equation}
It is numerically identical to the instability discovered by \citet{bell04}, but as we discuss in \cref{subBell}, its derivation is based on a slightly different argument.

\Cref{figKmax} compares these approximations with numerical solutions for the fastest-growing modes as a function of beam velocity $v_b$ (in the electron frame) and density $n_b$. Both predictions for $\overline k_\parallel$ are fairly accurate even for a relatively large beam density, provided that $v_b$ is large enough that the short- and large-wavelength modes are clearly separated (\cref{figKmax}a). However, the approximations for $\overline \gamma$ become increasingly misleading as the beam gets denser. Whereas Gary's $\overline \gamma_{\mathrm{RHI}}$ remains accurate within 10~\% for $n_b<0.35\ n_e$ in the case shown in \cref{figKmax}b, $\overline \gamma_{\mathrm{NRI}}$ overestimates the growth rate of the NRI by at least 50~\% for any value of $n_b$ in that case. But even though the quantitative usefulness of $\overline \gamma_{\mathrm{NRI}}$ may be limited at this moderate value of $v_b/v_A$, the important qualitative conclusion that Winske and Leroy drew remains valid: As the beam becomes denser, the growth rate of the left-handed NRI (or G/E) increases faster than that of the right-handed RHI (or F/A) and eventually exceeds it.

We try to summarise this discussion in \cref{tabVeryConfusing}, in which we also refer to previous derivations in the space-physics and astrophysical literature where they exist.

Let us take a step back from solving algebraic curves on $\mathbb C \times \mathbb R$ and consider the physical meaning of these modes. As noted before, the F/A and A/F instabilities couple the fast magnetosonic mode in one medium and the shear-Alfv\'en mode in the other medium, both in the long-wavelength regime in which using these MHD terms is justifiable. Thus both modes interact strongly with the perpendicular velocity components of the ion populations. A small perturbation in the perpendicular velocity of one ion species, e.g.\ a slight inhomogeneity of the background-ion gyrophase distribution, excites a propagating electromagnetic field that leads to oscillations of the perpendicular velocity not only of nearby background ions, but also of nearby beam ions. The perturbation of the perpendicular electric current is therefore larger than it would be without the beam and the response of the electromagnetic field is larger than the initial driving impulse. Provided that the beam ions can carry a wave with the same wavelength and Doppler-shifted frequency as the background ions, both ion species oscillate in resonance with each other and the A/F instability grows. 

\begin{table}
\begin{tabular}{lcr}
Name(s) & Wavenumber & Sample references\\
\toprule[1.1pt]
Beam-resonant modes & $0<k_\parallel\delta$ &\\
\hline
\quad E/G & $1<k_\parallel\delta$ &\\
\quad Electr./ion whistler & & \hskip-.6cm\cite{akimoto87}\\
\hline
\quad F/A & $0<k_\parallel\delta<1$ &\\
\quad RHI & & \hskip-.6cm\cite{gary78}\\
\quad CR-Alfv\'en instab. & & \hskip-.6cm\cite{kulsrud69}\\
\midrule[.8pt]
NRI & $k_\parallel\delta<0$ & \hskip-.6cm\cite{winske84}\\
Bell instability$^\ast$ & &\hskip-.6cm\cite{bell04}\\
\hline
\quad A/F & $-1<k_\parallel\delta<0$ &\\
\hline
\quad G/E & $k_\parallel\delta<-1$
\end{tabular}
\caption{Electromagnetic instabilities at $\vec k\times\vec B_0=0$ for a parallel ion-beam configuration in the cold hybrid model. $^\ast$See \cref{subBell}.}
\label{tabVeryConfusing}
\end{table}

\subsection{Warm Maxwellian beams}

In the previous section, we obtained exact solutions for the zero-temperature limit of both background and beam ions. Under the more general assumption that the velocity distributions of both ion species, as well as electrons, are Maxwellian with thermal velocity $v_{\mathrm{th},\alpha}$, the dispersion relation for parallel instabilities becomes
\begin{multline}
	\left( \omega^2 - k_\parallel^2~c^2 \right)~\frac{n_e}{\omega_p^2} + n_0~\zeta_0^{(0)}~Z\left(\zeta_0^{(1)}\right) - n_e~\zeta_e^{(0)}~Z\left(\zeta_e^{(1)}\right)\\ + n_b~\frac{q_b^2}{m_b}~\zeta_b^{(0)}~Z\left(\zeta_b^{(1)}\right) = 0,
\end{multline}
where $Z$ is the plasma dispersion function investigated by Fried and Conte and $\zeta_\alpha^{(n)} = (\omega-v_\alpha k_\parallel+n v_\alpha)/(k_\parallel v_{\mathrm{th},\alpha})$ with $\alpha\in\{0,b,e\}$. We employ \textsc{Whamp} \citep{roennmark1982} to numerically solve for the growth rate using Newton's method with a Pad\'e-approximated $Z$.

\begin{figure}
\centering
\includegraphics[scale=.95]{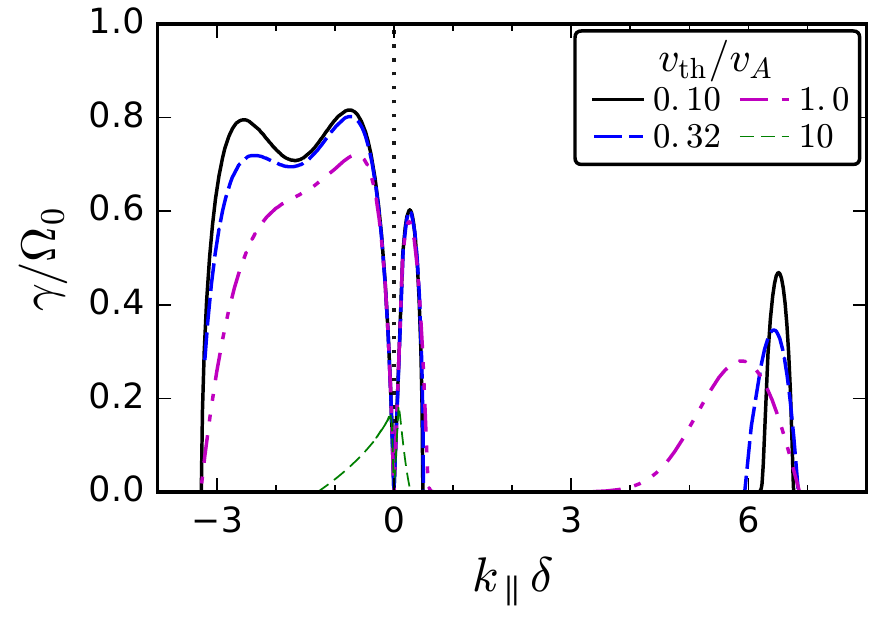}
\caption{Growth rates for parallel ion-beam instabilities for various ion temperatures, assuming a Maxwellian beam with $V_b=10\,v_A$ and $n_b=0.5\,n_0$ in a Maxwellian background plasma with $T_e = T_i$.}
\label{figWarmDR}
\end{figure}

For a parameter set at which all four modes are visibly separated, \cref{figWarmDR} shows that increasing the temperatures of both ion species mainly influences the short-wavelength instabilities, E/G and G/E. With the ion population spreading out in parallel-velocity space, the range of unstable wavelengths is extending towards smaller values of $|k_\parallel|$, while the largest growth rate decreases in proportion to the phase-space density of resonating ions. As low-frequency instabilities, the A/F and F/A instabilities are only slightly diminished by a rising ion temperature as long as the thermal velocity is smaller than the beam velocity. When the thermal ion velocity becomes comparable to the drift velocity of the beam, both modes grow at only a quarter of the original rate, as the thin dashed line in \cref{figWarmDR} shows for $v_{\mathrm{th},0}=v_{\mathrm{th},b}=V_b=10\,v_A$. Neither short-wavelength mode is unstable at this temperature.

As we have already pointed out, the two short-wavelength modes couple electron-whistler waves to the gyromotion of either the beam ions (for $k_\parallel > 0$) or the background ions (for $k_\parallel < 0$). Hence increasing only the beam-ion temperature while keeping the background temperature small affects only the E/G peak at $k_\parallel\delta>1$ and leaves the growth rates of all other modes essentially unchanged. This is the regime in which \citet{bell04} operated, although he neglected the background ions and instead considered the left tail ($v_\parallel<0$ in the electron frame) of an almost infinitely hot beam-ion distribution. Conversely, increasing only the background-ion temperature reduces the growth of the G/E mode at $k_\parallel\delta<-1$. Changing the ion and electron temperatures independently of each other, one finds that the effect of $v_{\mathrm{th},e}$ on all four ion-beam modes parallel to the magnetic field is negligible, as pointed out above. The electron temperature becomes important for oblique and perpendicular modes, however \citep[e.g.][]{sauer99,zweibel10}.

In conclusion, the ion temperature has a negligible effect on the F/A and A/F growth rates as long as $\beta\leq1$. On the other hand, using the cold-plasma dispersion relation for the E/G and G/E modes will overestimate these short-wavelength modes after the ion populations (beam and background, respectively) have been heated to even moderate temperatures. Because wave growth and ion heating are concomitant processes, their dynamic interaction is best investigated in numerical hybrid simulations.

\section{Growth-stage simulations}
\label{secGrowth}

\subsection{Numerical set-up}

Since we are interested in the dynamics on scales of tenths to hundreds of ion inertial lengths, we use a hybrid code which models ions by the particle-in-cell method and electrons as an inertia-less MHD fluid. The code was originally developed in the 1980s \citep{winske84} and, being continuously extended, has been widely used in the space-physics community \citep[e.g.][]{quest88,akimoto93} and more recently in the context of laboratory astrophysics \citep{clark13,weidl16,heuer18}. It evolves the magnetic field according to a discretised Faraday's equation, employing subcycling when necessary, and derives the electric field from
\begin{equation}
	\vec E = \frac{\left(\nabla \times \vec B\right) \times \vec B}{e~n_e} - \frac{\vec U_i \times \vec B}{c} - \frac{\nabla p_e}{e~n_e}.
\end{equation}

In the fiducial configuration, the electromagnetic fields are evolved on a two-dimensional grid of $1024\times160$ grid cells, corresponding to a physical size of $L_x \times L_y = 128\,\delta\times20\,\delta$. For the beam velocities we use, this resolution captures both of the high-$|k_\parallel|$ modes derived above while several wavelengths of the low-$|k_\parallel|$ modes fit into these dimensions. When necessary, the perpendicular size of the domain is increased such that it always measures at least two beam-ion gyroradii. The background ion species is modelled with 100 quasiparticles (or quions) per grid cell. These background quions are initially distributed homogeneously over the entire simulation domain with a Maxwellian velocity distribution corresponding to $T_{0,i} = 1.5$ eV, assuming the proton mass for the ions and $c/v_A = 1739$. With a homogeneous magnetic mean-field $\vec B_0$ pointing from now on along the $x$ direction, the plasma beta is $\beta = 2 n_0 T_{0,i}/B_0^2 = 0.01$. The electron fluid density $n_e(\vec x)$ is identical to the total ion density everywhere in order to conserve charge neutrality, with an initial temperature $T_{0,e} = T_{0,i}$ and a polytropic electron-pressure model $p_e(\vec x) \propto n_e(\vec x)^{5/3}$. We have found no remarkable difference in comparison runs with an isothermal electron model. We introduce no artificial dissipation at high $|k_\parallel|$ (beyond numerical diffusion) in order to facilitate comparison with the linear calculations in the previous section.

To obtain the mean ion velocity $\vec U_i$, we average over the velocities of both background and beam quions in each grid cell. The beam-ion population is also initially homogeneously distributed, with each quion representing the same number of protons as each background quion. For each beam quion, we determine the starting velocity as the sum of an isotropic Maxwellian component with temperature $T_{0,i}$ and a drift component $\vec v_d$, which points on average into the positive $x$ direction -- a small random deviation for each quion of up to $1^\circ$ models a possible distant point source of the beam. The time step of every simulation is chosen to ensure that the Courant--Friedrichs--Lewy criterion is fulfilled at all times.

\begin{figure}
\centering
\includegraphics[scale=.95]{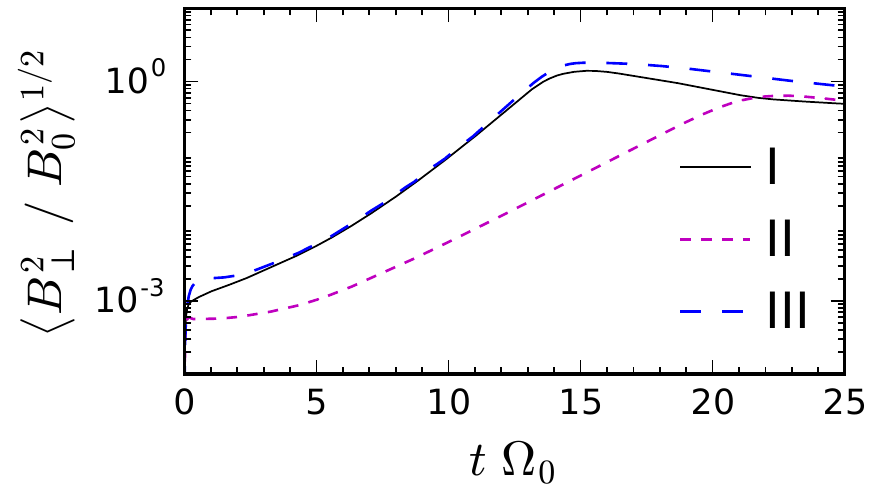}
\caption{Growth of the perpendicular magnetic field averaged over the entire domain for Runs I (solid), II (magenta short-dashed), and III (blue long-dashed).}
\label{figLinGrowthRuns}
\end{figure}

The three parameter sets that we describe in greater detail below are chosen to yield dominance of the left-handed A/F instability (Run~I), dominance of the right-handed F/A instability (Run~II), and a balanced mix of left- and right-handed polarisation (Run~III). \Cref{figLinGrowthRuns} compares the growth of the magnetic field in each run. We first outline the linear stage of each run and compare the field spectra before saturation to the theoretical framework summarised in the previous section; the non-linear saturation of each simulation follows in \cref{secSaturation}.

\subsection{Run I: Medium beam density}

We first present the results of a simulation with relative beam density $n_b/n_0 = 0.5$ and relative beam drift $V_b = |\vec v_d| = 10.0~v_A$. The linear growth rates in \cref{figWarmDR} are based on these conditions, assuming isotropic Maxwellian distributions for background and beam ions. Although the two modes with negative helicity (G/E and A/F) can be distinguished at low background-ion temperatures, the peaks are already very close together and become impossible to tell apart at $v_\mathrm{th}\sim v_A$. We will therefore use the non-specific term `NRI' to refer to a spectrum with significant power in the wavenumber range $-3<k_\parallel\delta<0$.

\begin{figure}
\centering
\includegraphics[scale=.95]{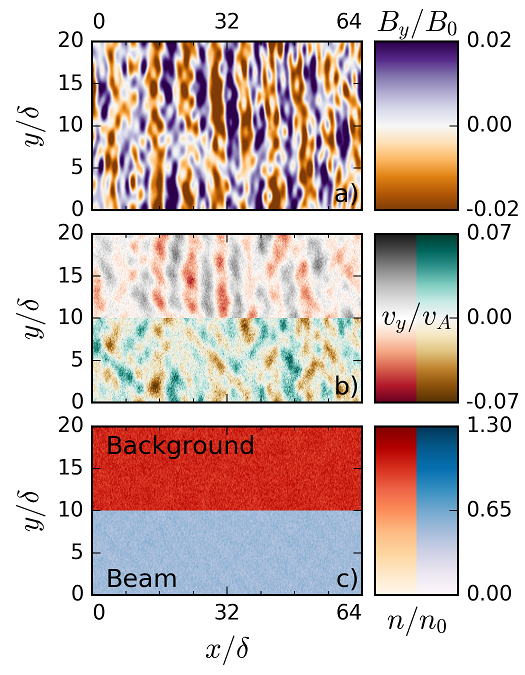}
\caption{Run~I: Profiles at $t\,\Omega_0=7.5$ of (a) the magnetic-field component $B_y$, (b) the velocity component $v_y$, (c) the plasma density $n$. The plots for $v_y$ and $n$ show the background plasma for $y>10\,\delta$ and the beam plasma for $y<10\,\delta$.}
\label{figProfileMed1}
\end{figure}
\begin{figure}
\centering
\includegraphics[scale=0.95]{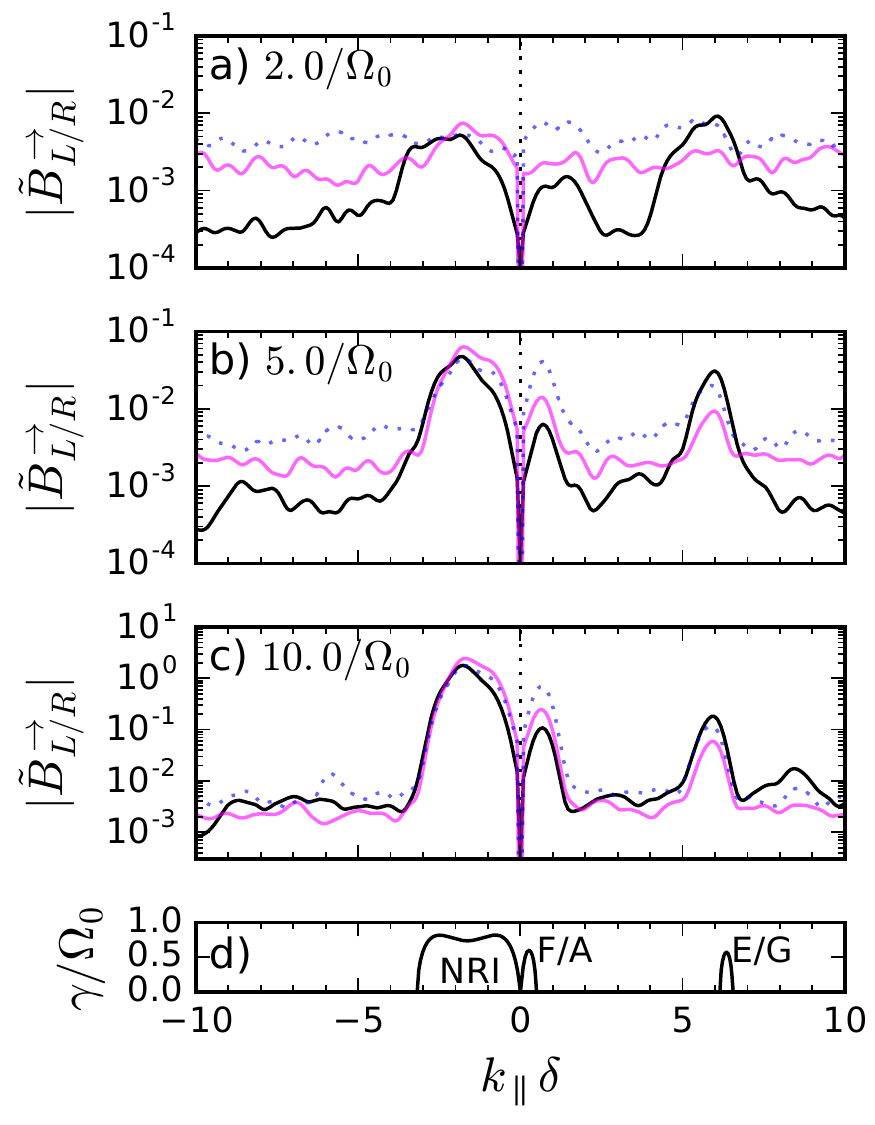}
\caption{Run~I: a--c) Helicity-decomposed power spectra of the spatial Fourier transforms of the perpendicular magnetic fields $B_y \pm i B_z$ (black) and the perpendicular velocity of background ions (magenta) and beam ions (blue dots) at different times, smoothed with a Gaussian filter with $\sigma=0.25\,\delta^{-1}$. d) Linear growth rates for cold ion beams. For rightwards-propagating waves, $k_\parallel<0$ corresponds to left-handed circular polarisation (G/E and A/F or NRI), $k_\parallel>0$ to right-handed polarisation (F/A and E/G).}
\label{figSpectraMed1}
\end{figure}

\Cref{figProfileMed1}a shows that the perpendicular magnetic field is soon dominated by quickly growing parallel Alfv\'en waves. Having become visible in the magnetic field, this structure slowly begins to dominate the ion-velocity profiles, too. After five inverse gyrofrequencies, the background plasma is still barely perturbed by the beam, but at $t\,\Omega_0 = 7.5$, its perpendicular velocity has adopted a spatial structure similar to the magnetic field. Meanwhile, a mix of oblique and parallel modes is visible in the profile of the perpendicular velocity of the beam. We expect this from linear theory, which predicts for these beam parameters that the growth rate of the NRI decreases only very slowly as one increases the angle between the direction of the wave and the $x$ axis, by less than $2\%$ up to about $30^\circ$. Nevertheless, since the linearly incompressible, parallel Alfv\'en waves dominate, the plasma densities of both species remain largely homogeneous.

Although \cref{figProfileMed1} makes it clear that each plasma component is dominated by modes with a wavelength of $\lambda\approx3\,\delta$, it gives no hint as to which way the waves are polarised. This question is resolved in Fig.~\ref{figSpectraMed1}, in which we show the helicity decompositions of the perpendicular magnetic field and of the background velocity. This technique \citep[e.g.][]{weidl16} allows us to distinguish parallel modes with $k_\parallel < 0$, such as the NRI, from modes with positive helicity, like the F/A and E/G mode, for an easier comparison with linear theory. All growing waves propagate to the right in the simulation frame; hence $\omega/k_\parallel>0$ and negative wave vectors correspond to left-handed circular polarisation and \textit{vice versa}.

Two inverse gyrofrequencies after the beam has begun propagating, the perpendicular velocity spectrum of the background-ions is still rather flat, with only a minor hump at negative wavenumbers (magenta line in \cref{figSpectraMed1}a). Not being subject to particle shot noise, the magnetic field (black) evolves a clear structure first, exhibiting a global maximum at $k_\parallel \delta \approx +6$, the position of the right-handed E/G instability. The other right-handed instability, the F/A mode, is also already visible in the magnetic spectrum as a relatively broad peak at $k_\parallel \delta \approx +1$. Although cold-plasma theory predicts equal growth rates for both modes (\cref{figSpectraMed1}d), the amount of self-organisation required for the longer-wavelength, lower-frequency F/A instability takes more time. Any theory of electromagnetic plasma waves can only be a vague approximation at timescales smaller than one gyroperiod or $2\pi / \Omega_0$.

Hence it takes almost as long, or five inverse gyrofrequencies, for the left-handed modes to catch up with the E/G peak (\cref{figSpectraMed1}b). With the parameters of Run~I, the G/E and A/F modes merge into one broad-spectrum instability that we will collectively call the NRI after \citet{winske84}, although we emphasise again that it is driven by a resonant interaction between collective plasma excitations and the gyromotion of background ions. Hence it appears first in the spectrum of the perpendicular background-ion velocity. According to the linear theory for cold ion beams, this instability should grow faster than either right-handed mode by about $30\,\%$. At $t=10/\Omega_0$, the helicity-decomposed spectrum (\cref{figSpectraMed1}c) matches this prediction extremely well. Consequently, the NRI wavelength $\lambda\approx3\,\delta$ dominates both the magnetic-field and the background-velocity profiles in \cref{figProfileMed1}a and b and continues to do so throughout the linear-growth phase. From this point onwards, neither the growth rate nor the dominant wavelength change significantly until the NRI saturates at $t\,\Omega_0 \approx 14$.

Going back to \cref{figLinGrowthRuns}, we note that the transition from the E/G- to the NRI-dominated stage is also visible in the evolution of the perpendicular magnetic-field energy as a change of slope around $t\,\Omega_0\approx6$. During the interval $2<t\,\Omega_0<5$, while the E/G mode is still dominating, the root-mean-square of $B_\perp$ grows at a rate of approximately $0.37\,\Omega_0$. When the large-scale structure of the NRI has eventually formed, the field amplification rate almost doubles to $0.66\,\Omega_0$ (averaged over $8<t\,\Omega_0<13$). These rates are consistent with the linear-theory predictions for $v_{\mathrm{th}}=0.32\,v_A$.

\begin{figure}
\centering
\includegraphics[scale=.95]{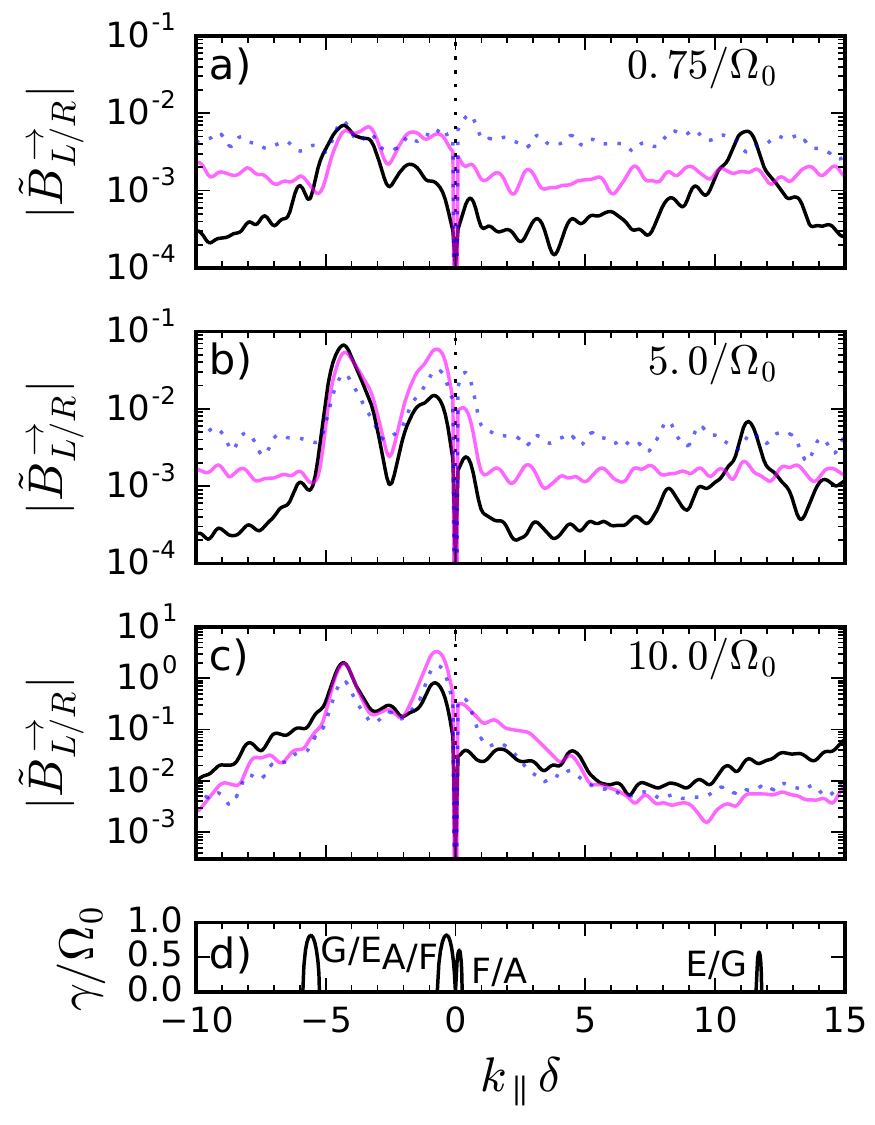}
\caption{As \cref{figSpectraMed1}, but for $V_b=17.8~v_A$.}
\label{figSpectraMed2}
\end{figure}

To prove beyond doubt that the NRI consists of two separate modes, we plot in \cref{figSpectraMed2} the helicity-decomposed spectra for a hybrid run with a higher beam velocity, but at the same time intervals as above. For $V_b=17.8\,v_A$, linear theory predicts that the G/E and A/F modes split up into clearly distinct peaks. In the corresponding hybrid simulation, the two whistler-type modes (the left-handed G/E or Bell mode and the right-handed E/G mode) grow similarly fast at early times, at least in the magnetic field (black line in \cref{figSpectraMed2}a). All four modes are easily identifiable after five inverse gyrofrequencies in \cref{figSpectraMed2}b, although the background ions do not get very excited by the E/G peak at a frequency of $\omega\approx140\,\Omega_0$. As the system finally approaches saturation in \cref{figSpectraMed2}c, the E/G mode has disappeared completely because of an increased beam-ion temperature. While remnants of the F/A mode are still visible in the velocity spectra of both species, the left-handed A/F and G/E modes now unambiguously dominate all spectra.

In summary, the linear predictions for a strong left-handed polarisation match the simulations of this regime very well, but only once a common structure shared by the magnetic field and the velocity fields has emerged; compare how the spectra in each subplot of \cref{figSpectraMed1} converge over time. First, the right-handed polarisation of the E/G mode dominates the perpendicular magnetic field in an intermediate stage, during which energy is primarily transferred from the gyromotion of the beam ions to the field and the electrons. The background ions at this point do not yet reflect the E/G mode in their velocity profile since this wave, an electron whistler from their perspective, has a frequency $\omega\gtrsim60\,\Omega_0$, far too large to influence their trajectory effectively at low amplitude. Yet once the background ions have had time to shape the magnetic field, resulting in the growth of the NRI, both ion species begin to synchronise their perpendicular velocity profiles into a shared, predominantly left-handed structure (in the background frame). We will show below that this synchronisation is already a first step towards saturation.

\subsection{Run II: Low beam density}

In the second run, the relative beam density has been reduced to $n_b/n_0 = 0.25$ and the drift velocity to $V_b = 5.6\,v_A$. Unsurprisingly the magnetic field grows significantly more slowly and thus takes longer to reach a saturated level (Fig.~\ref{figLinGrowthRuns}). 

\begin{figure}
\centering
\includegraphics[scale=0.95]{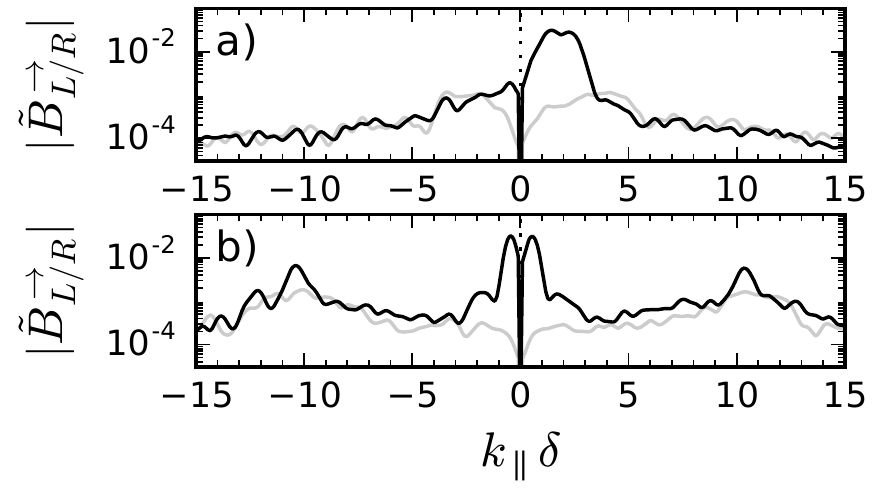}
\caption{Power spectra of the spatial Fourier transforms of the perpendicular magnetic fields $B_y \pm i B_z$ at $t\,\Omega_0=0.5$ (light grey) and $10.0$ (solid black) for a) the right-hand dominated Run~II and b) the mixed-turbulence Run~III.}
\label{figSpectraLoHi}
\end{figure}

Similar to the previous run, the early stage of Run~II shows relatively fast growth of the right-handed E/G mode in the whistler range, while at low frequencies the magnetic-field growth remains balanced between both polarisations. The helicity-decomposed spectrum of the magnetic field at $t\,\Omega_0=0.5$ (grey line in \cref{figSpectraLoHi}a) thus has a weak global maximum at $k_\parallel\delta=+4$. As time progresses, longer-wavelength fluctuations are eventually able to organise coherently. Meanwhile the saturation threshold of short-wavelength fluctuations is lowered by heating of the ion populations, analogous to the previous case. But unlike in Run~I, which gradually became dominated by left-handed modes, the lower beam parameters of Run~II result in fast growth of the right-handed F/A instability at $k_\parallel\delta\approx+1$. Hence, after eight inverse gyroperiods (black line in \cref{figSpectraLoHi}a) the helicity of the magnetic fieldlines is still predominantly positive, i.e.\ the polarisation of the strongest waves has remained right-handed.

\begin{figure}
\centering
\includegraphics[scale=.95]{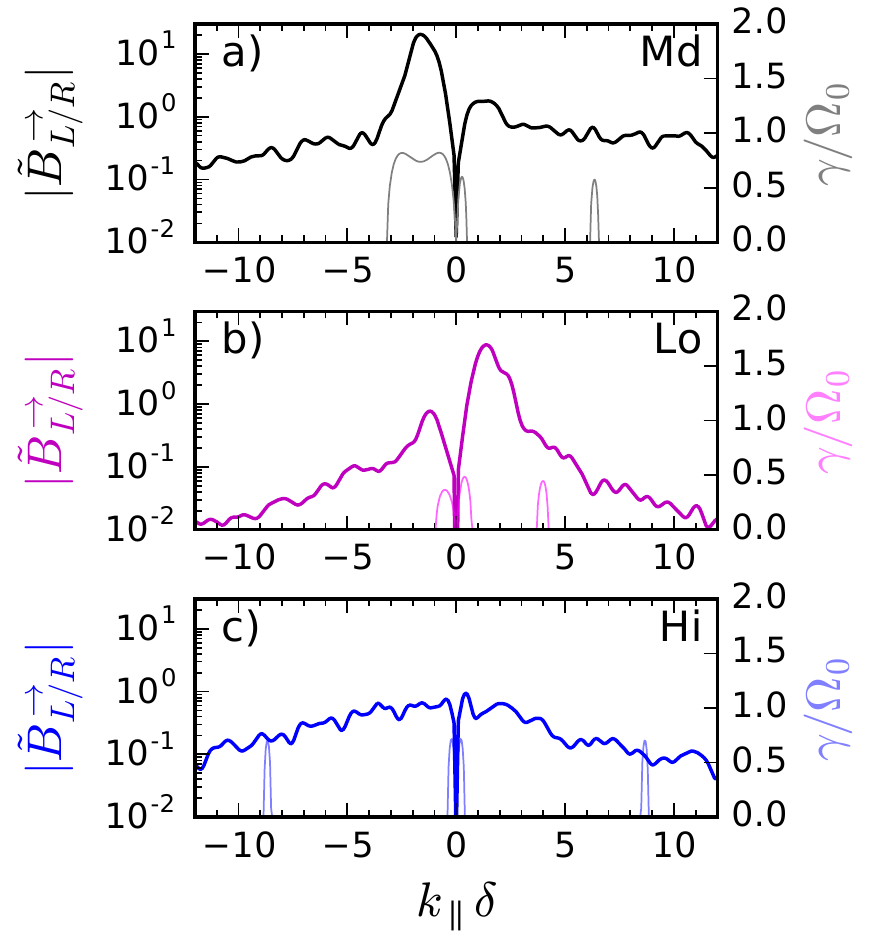}
\caption{Comparison of helicity-decomposed spectra of the perpendicular magnetic field (left axis) and the growth rates predicted by linear cold-plasma theory (right axis) for a)~Run~I at $t\,\Omega_0=14.0$ b)~Run~II at $t\,\Omega_0=21.0$, c)~Run~III at $t\,\Omega_0=14.0$.}
\label{figMLHspectra}
\end{figure}

The growth during the interval $5<t\,\Omega_0<20$ is best fit by $B_\perp \propto e^{\gamma t}$ with $\gamma \approx 0.4\,\Omega_0$. This rate is in good agreement with linear-theory results for this parameter choice, shown at the bottom of Fig.~\ref{figMLHspectra}b. Also plotted in that graph is the helicity decomposition at the time of saturation ($t\,\Omega_0=21$), when the root-mean-square perpendicular magnetic field reaches its maximum. The peak spectral power at this time is located at $k_\parallel\delta \approx +1.3$, between the theoretical F/A wave number of $k_\parallel\delta \approx +0.4$ and the theoretical E/G peak at $k_\parallel\delta \approx +4.0$. The latter mode has disappeared all but completely; it has left behind only a small bump beside the large peak of the F/A instability.

Thus the helicity (and polarisation) of the dominant wave-type remains positive throughout the linear-growth stage of Run~II. However, as in Run~I, field amplification shifts from short wavelengths to long wavelengths.

\subsection{Run III: High beam density}

In the third run, we increase the relative beam density to $n_b = n_0$ and the drift velocity to $V_b = 17.8 v_A$. With the beam and background density being equal, the distinction between both ion species becomes almost arbitrary. Linear cold-plasma theory predicts that all four modes grow equally fast (bottom of \cref{figMLHspectra}c). In the simulation, we observe once more that the whistler-type instabilities grow fastest initially, as shown by the grey line in \cref{figSpectraLoHi}b for $t\,\Omega_0=0.5$.

However, their further development depends strongly on the frame in which the simulation is set: Numerical dispersion effects tend to damp high-$k_\parallel$ modes with a large phase velocity; a run set in the background-ion frame thus tends to develop an asymmetry favouring the G/E mode ($\omega/k_\parallel\approx0.1\,v_A$) over the E/G mode ($\omega/k_\parallel\approx6\,v_A$). Running an otherwise identical setup in the zero-current frame, in which both ion species start with opposite drifts $V_x=\pm8.9\,v_A$, or increasing the spatial resolution to minimise numerical resistivity, we obtain a perfectly balanced spectrum for the whistler modes (as shown in \cref{figSpectraLoHi}b for the symmetric setup).

Entirely unaffected by the choice of reference frame (or any other detail of the initial conditions), long wavelengths dominate when either simulation reaches its peak magnetic-field strength at $t\,\Omega_0\approx14$. As the perturbations reach nonlinear amplitudes, wave packets of different modes begin to interact with each other and decay into a turbulent cascade which favours neither helicity. Simultaneously the increasing ion temperature leads to the saturation of both whistler-type instabilities. In \cref{figMLHspectra}c we show that, shortly before saturation is reached (as defined below), Run~III exhibits a balanced turbulence spectrum in both helicities and wave-wave interactions have smoothed out any spectral peak. Even though the total beam energy is larger than in Run~I by more than a factor of 6 and larger than in Run~II by a factor of 40, the most energetic mode of the magnetic field in Run~III contains only about 1\% of the spectral energy in the F/A peak in Run~II.

\begin{figure}
\centering
\includegraphics[scale=.95]{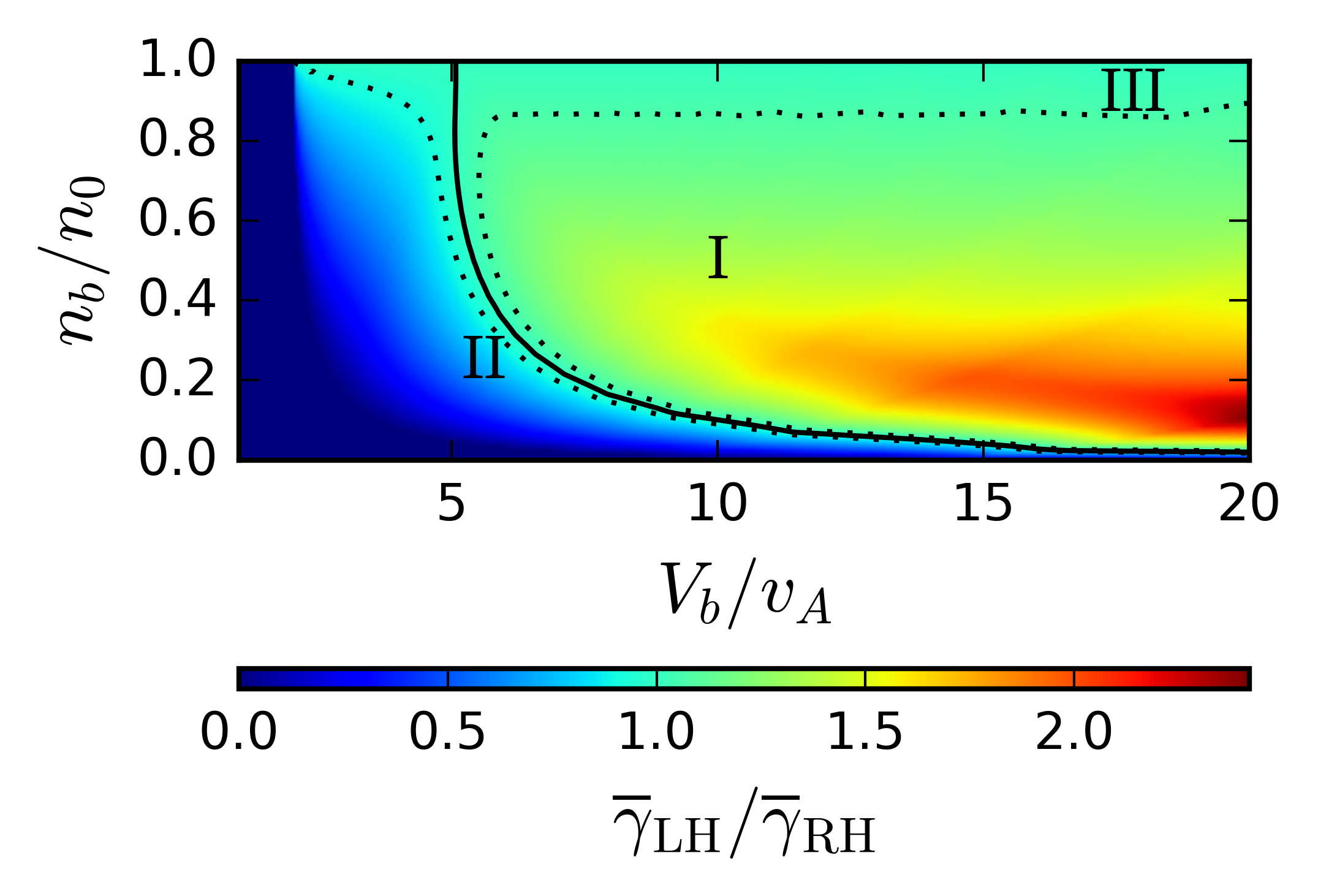}
\caption{Positions of Runs~I, II, and III in $V_b$-$n_b$ parameter space and contour of equal growth rates (solid) for left-handed modes ($\overline{\gamma}_{\mathrm{LH}}$) and right-handed modes ($\overline{\gamma}_{\mathrm{RH}}$) in the background frame for $q_b=m_b=1$. Dotted lines mark ratios of $\overline{\gamma}_{\mathrm{LH}}/\overline{\gamma}_{\mathrm{RH}}=0.93,1.07$ respectively.}
\label{figRatioLR}
\end{figure}

This mixed case constitutes the third possible regime of parallel electromagnetic waves created by a streaming ion beam. Depending on the density and relative velocity of the beam, the resulting wave spectrum in the linear stage can be either predominantly left-handed (the NRI regime that Bell considered), right-handed (the beam gyroresonance regime), or of mixed polarisation and turbulent. The exact ratio of A/F and F/A growth, derived from eqn.~\eqref{eqnBeamDR}, is given in \cref{figRatioLR}. 

\section{Saturation in hybrid simulations}
\label{secSaturation}

\subsection{Run I: Medium beam density}

\begin{figure}
\centering
\includegraphics[scale=.95]{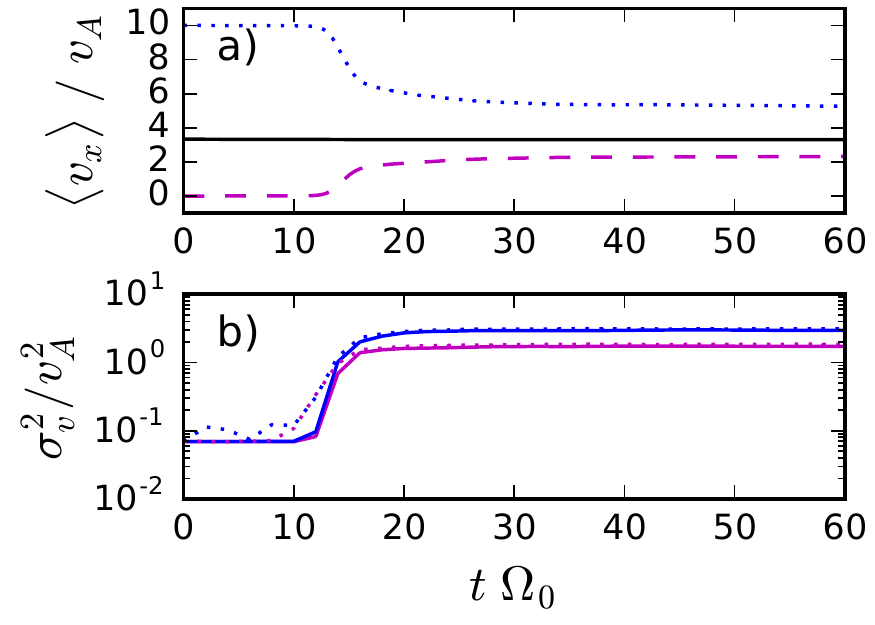}
\caption{Run I: Evolution of the a) average $x$-velocity in units of $v_A$ of the background ions (magenta dashed), the beam ions (blue dotted), and all ions (black solid), b) parallel (solid) and perpendicular (dotted) velocity dispersion, with respect to the $x$ axis, in units of $v_A^2$ for background and beam ions.}
\label{figSatEvol}
\end{figure}

Returning to the left-hand dominated Run I ($V_b=10\,v_A, n_b = 0.5\,n_0$), we now move on to analyse how the NRI saturates. As we are about to see, this process involves both resonant scattering and turbulent heating of ions. Hence the most informative plots, together with the perpendicular magnetic energy shown in \cref{figLinGrowthRuns}, are of the evolution of the bulk velocity along the magnetic-field direction and the velocity dispersion of both ion species (\cref{figSatEvol}). The three plots imply a clear sequence of events: a rapid increase of the perpendicular temperatures that is first visible at $t\,\Omega_0\approx10$, followed by an equal increase in the parallel velocity dispersion as both bulk velocities start to converge towards the average ion velocity ($t\,\Omega_0\approx12$), and finally a peak in the magnetic energy at $t\,\Omega_0\approx14$, which slowly drops back towards an equilibrium value.

\begin{figure}
\centering
\includegraphics[scale=.95]{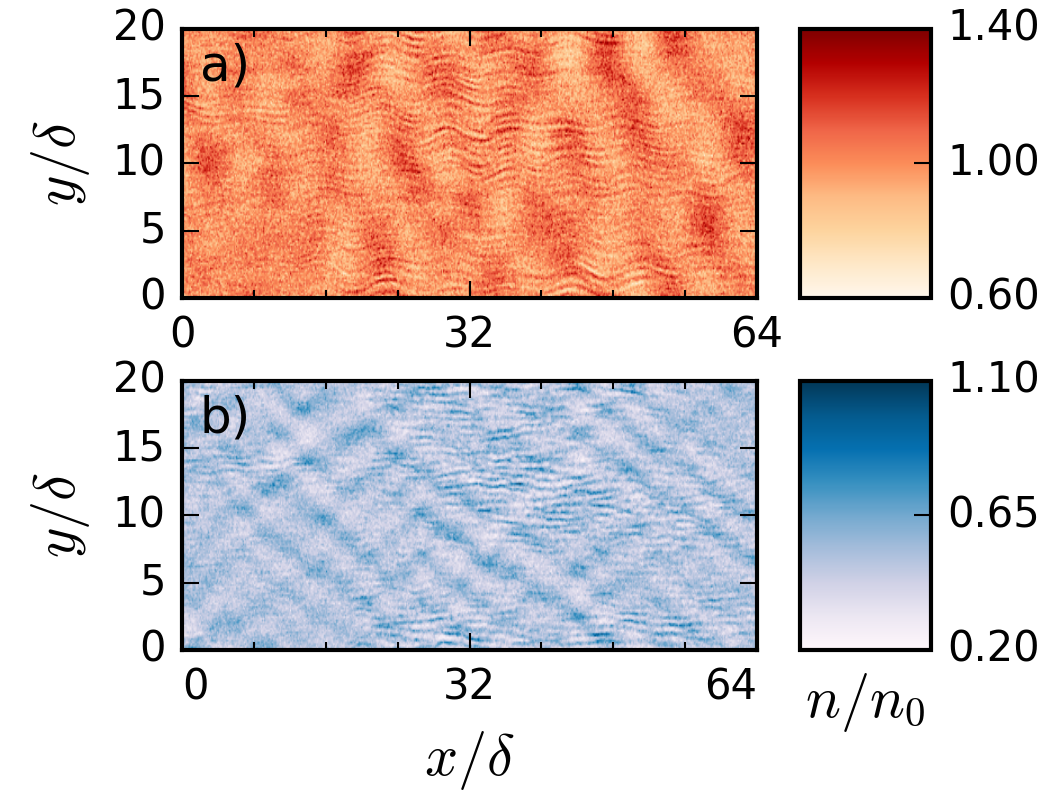}
\caption{Run I: Density profiles of background ions (a) and beam ions (b) at $t\,\Omega_0=11$.}
\label{figDProfile}
\end{figure}

As an early indicator of saturation setting in, the density profiles of both ion populations develop compressive features, albeit to different degrees. \Cref{figDProfile}a shows that, for $t\,\Omega_0=11.0$, the density fluctuations in the background medium are still primarily longitudinal along the $x$ axis. From the perspective of the background ions,  the interaction with the magnetic field can still be approximated as parallel Alfv\'en waves of nonlinear amplitude. Contrariwise, the structure of the compressions of the beam medium is dominated by strongly oblique modes, with the dominating wavevectors subtending about $50^\circ$ with the $x$ axis (\cref{figDProfile}b). The normalised compression of the beam $\langle\delta n_b\rangle_{\mathrm{RMS}} / n_b = 16.5\%$ is almost double that of the background ($8.6\,\%$) because the beam ions are excited in the pattern of a compressive fast magnetosonic mode.

When the perpendicular magnetic field $B_\perp$ grows larger than the mean-field $B_0$ (\cref{figLinGrowthRuns}), nonlinear effects truly take over: The almost isotropic magnetic field rapidly scatters the beam ions first perpendicularly and then reduces their averaged flow velocity along the mean-field direction from $V_b=10\,v_A$ to $5.3\,v_A$ (measured in the simulation frame). Simultaneously the background ions, with a negative $x$ velocity in the frame of the rightwards-propagating A/F waves, are accelerated to a velocity of $-1.0\,v_A$ in the centre-of-mass frame or $+2.3\,v_A$ in the simulation frame. Throughout this process the centre of mass of the combined ion population maintains its initial velocity $v_{CM} = n_0 v_{x,0} + n_b v_{x,b} \equiv 3.3\,v_A$ to an accuracy of better than one percent. In other words, the magnetic spectrum evolves in such a way that the momentum gained by acceleration of the background ions balances the momentum lost by backward scattering of the beam ions at all times.

\begin{figure}
\centering
\includegraphics[scale=.95]{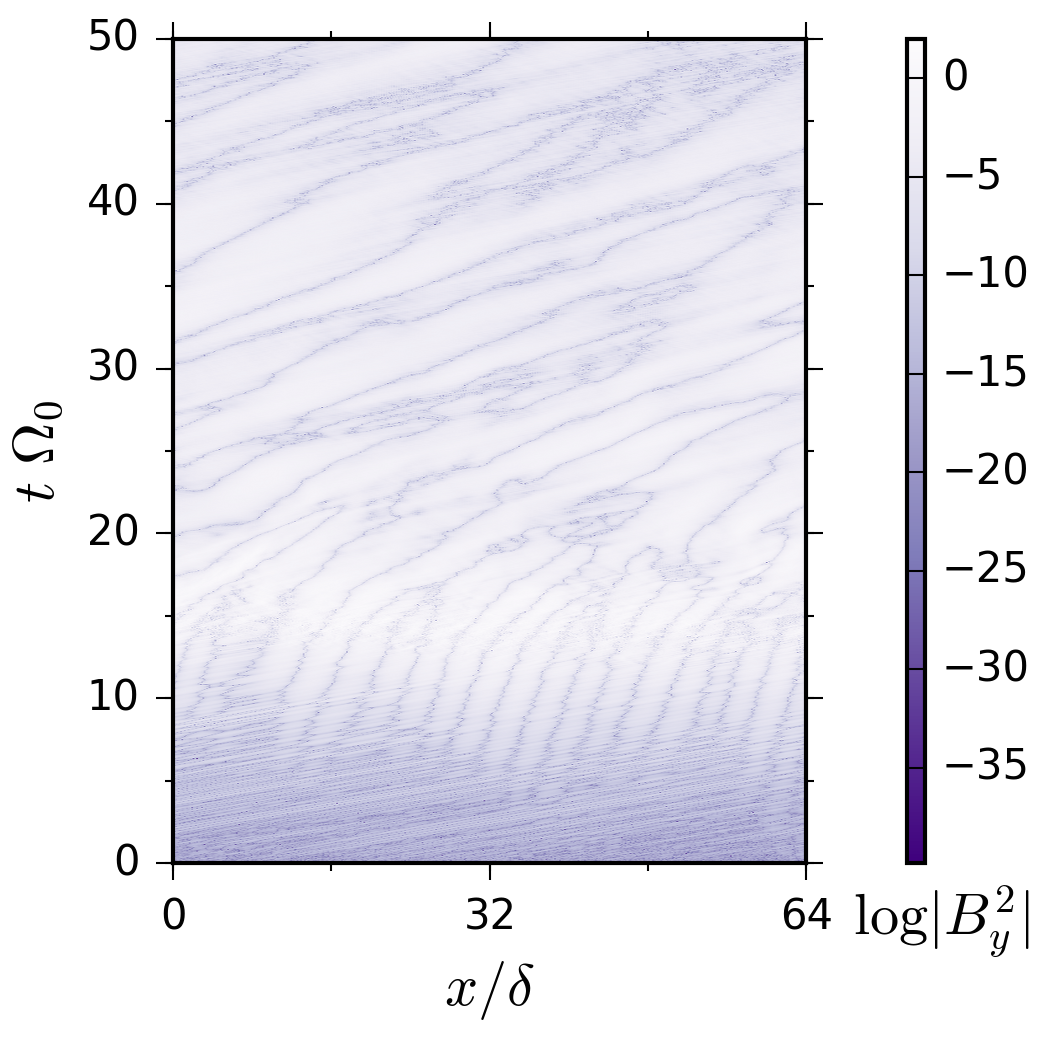}
\caption{Run I: Evolution of magnetic-field intensity $|B_y^2|$ along the line $y=10\,\delta$, plotted in the simulation frame.}
\label{figByProfile}
\end{figure}

As the velocity difference between both populations decreases, the phase speed of the A/F waves increases rapidly in the simulation frame while their wavelength fluctuates wildly (\cref{figByProfile}) . The process saturates once the velocity difference is so small that the rate at which the A/F instability grows becomes negligible. In fact, the linear cold-beam growth rate for negative-helicity modes vanishes entirely below a beam velocity of $v_b=1.5\,v_A$ in the electron frame or $V_b=2.2$ in the background frame. We showed in \cref{secTheory} that even the long-wavelength modes become sensitive to the temperature at super-Alfv\'enic thermal velocities; therefore the A/F mode saturates while the beam velocity is still slightly larger. Because of ion heating ($\sigma_v^2=1.8\,v_A^2$ for background and $3.0\,v_A^2$ for beam ions) and because of the strongly nonlinear field amplitude, the final velocity difference measures $3.0\,v_A$. The fastest-growing wavelength that is predicted for these parameters, $\lambda=16\,\delta$, agrees well with the magnetic turbulence profile at $t\,\Omega_0=30$.

\begin{figure*}
\centering
\includegraphics{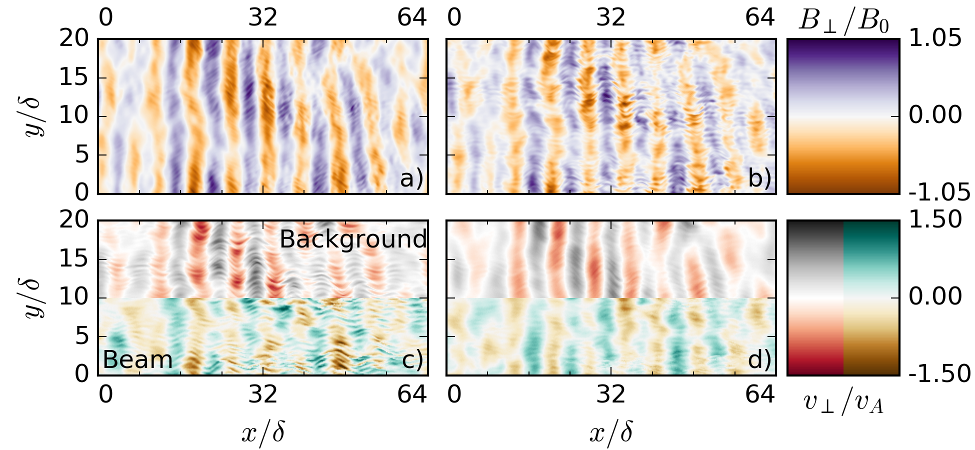}
\caption{Run I: Profile at $t\,\Omega_0 = 12.0$ of the perpendicular magnetic field a) $B_y$ and b) $B_z$, and the perpendicular velocity c) $v_y$ and d) $v_z$ for background ions (top half) and beam ions (bottom half).}
\label{figProfileMed2}
\end{figure*}

Going back to when saturation begins ($t\,\Omega_0 = 12.0$), \cref{figProfileMed2} shows the perpendicular components of the magnetic field. The profile is still dominated by the parallel A/F mode with $\lambda\approx6\ \delta$, but $B_y$ and $B_z$ have each grown comparable to $B_0$ by now. The same mode also shapes the perpendicular components of the bulk velocity of both ion species (\cref{figProfileMed2}c and d). This fact by itself is not surprising; we are observing parallel Alfv\'enic waves, which perturb both the velocity and the magnetic-field components in MHD theory, with a phase shift of either 0 or $\pi$ for shear-Alfv\'en waves and fast magnetosonic waves. Yet the phase shift between the velocity profiles of the background ions (which carry an Alfv\'en wave) and the beam ions (in whose frame the same wave becomes a fast magnetosonic wave) is clearly neither 0 nor $\pi$; instead the beam lags behind the background velocity by about $\pi/2$. Nor is either ion species completely in phase with the magnetic field.

\begin{figure}
\centering
\includegraphics[scale=.95]{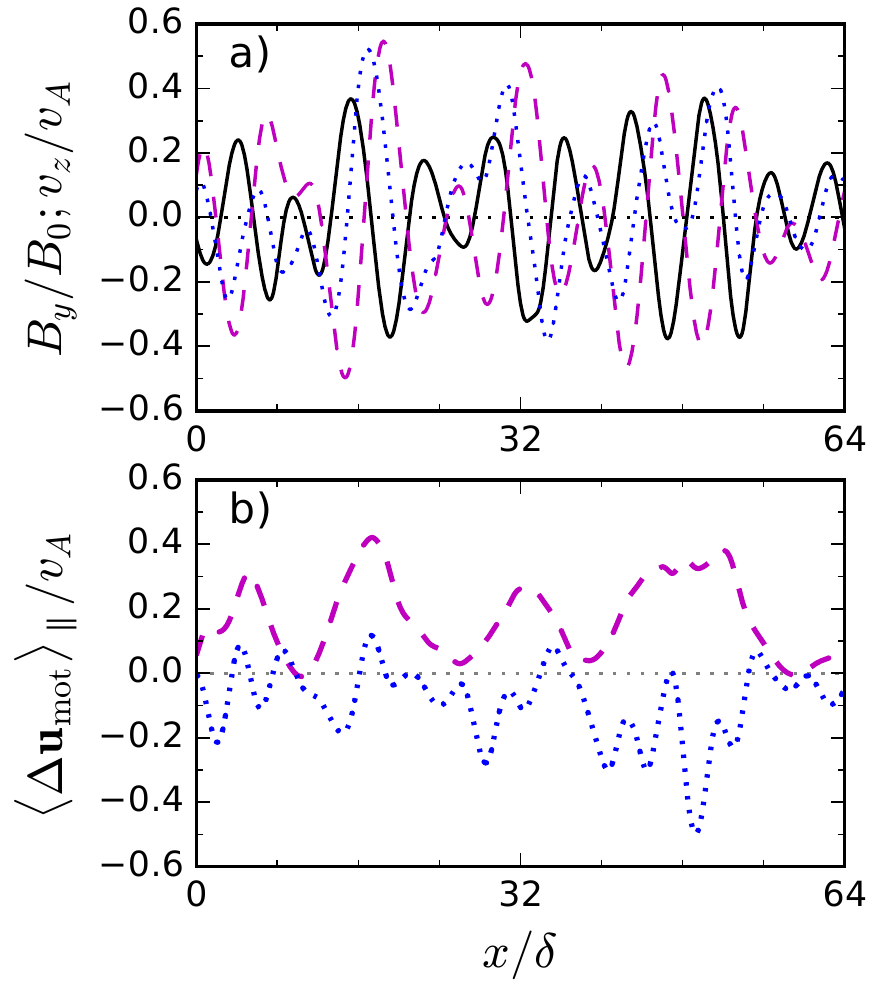}
\caption{Run I: a) Profiles of the $y$ component of the magnetic field (black solid), the $z$ components of the background-ion velocity (magenta dashed) and of the beam-ion velocity (blue dotted) at $t\,\Omega_0 = 12.0$ and along $y = 5\,\delta$. b) Net velocity change $\Delta \mathbf{u}_{\mathrm{mot}}$ due to the motional electric field parallel to the $x$ axis, integrated over $12<t\,\Omega_0<13$, for background (magenta dashed) and beam ions (blue dotted).}
\label{figPhaseMed}
\end{figure}

Since background ions and beam ions stream in opposite directions with respect to the strongly dominating A/F waves, a background ion that has obtained a large $v_z$ component in a region with a strong $B_y$ component is statistically more likely to fall behind on the left of the $B_y$ maximum, whereas a beam ion will overtake it and is more likely to be located on the right-hand flank of the peak. In a strict MHD picture the gyromotion of ions is ordered out, they follow the fieldlines instantaneously, and the velocity of both species is perfectly in-phase with the magnetic perturbations. From a more realistic ion-kinetic perspective retaining finite-Larmor-radius effects, the relative phases of the averaged perpendicular ion velocities must be affected by their parallel drift with respect to the waves. Hence, in \cref{figPhaseMed}a, the peaks invariably occur in the order $(B_y, v_{z,b}, v_{z,0})$ or (solid, dotted, dashed), with high values of $v_{z,b}$ predominating where $B_y>0$ and high values of $v_{z,0}$ where $B_y<0$. Integrated over one gyroperiod, this phase relation results in a net motional electric field that points in the positive $x$ direction for the background ions and in the negative $x$ direction for the beam ions (\cref{figPhaseMed}b).

\begin{figure}
\centering
\includegraphics[scale=.95]{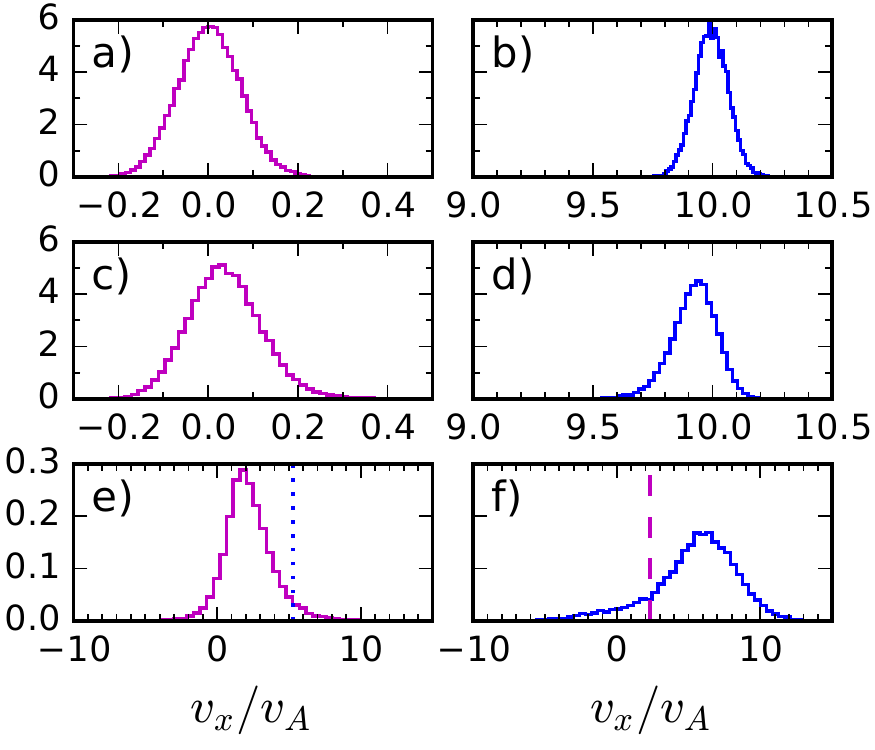}
\caption{Run I: Histograms of the parallel-velocity distribution for background (left) and beam ions (right) at (a,b) $t\,\Omega_0=10.0$, (c,d) $t\,\Omega_0=12.0$, (e,f) $t\,\Omega_0=30.0$, with the final bulk velocity of beam (blue dotted) and background (magenta dashed).}
\label{figHistoMed}
\end{figure}

Of course, no individual particle's gyrophase stays in any exact relation with the local magnetic field for the entire saturation stage, and as a result of phase mixing ions undergo incoherent scattering in all directions. Since this scattering of individual ions is still preferentially directed towards either the positive or the negative $x$ axis, the histograms of the global $v_x$ distribution for each species become increasingly asymmetric. \Cref{figHistoMed}a and b show that both background and beam ions maintain their isotropic Maxwellian distribution until shortly before saturation begins. At $t\,\Omega_0=12.0$, both the width and the skewness of each histogram start to increase. After the turbulence has reached a steady state, so have the histogram shapes (Figs.~\ref{figHistoMed}e,f) -- at least on the timescales we use in this analysis. By that point, 5.7\% of the beam population has been scattered to negative parallel velocities $v_x<0$ in the initial rest-frame of the background medium; 12.8\% of the beam population has a negative parallel velocity in the final rest-frame of the background ($v_x<2.3\,v_A$ in the simulation frame).

\begin{figure}
\centering
\includegraphics[scale=.95]{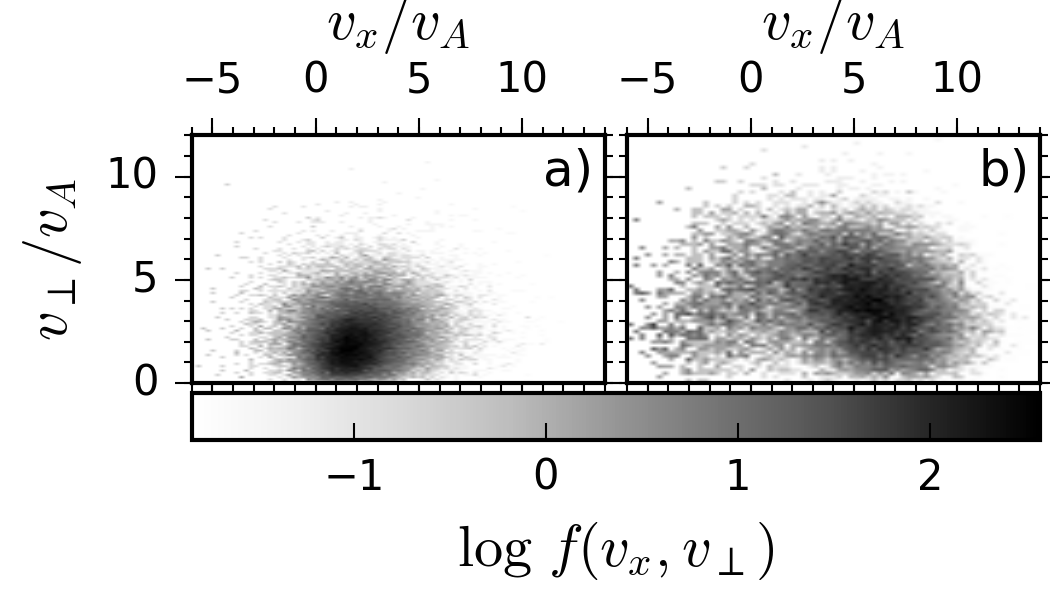}
\caption{Run I: Velocity-space density of (a) background and (b) beam ions in entire simulation domain at $t\,\Omega_0=30.0$, normalised to $\int f\, \mathrm d^2v_\perp\, \mathrm dv_x=1$.}
\label{figCrescentMed}
\end{figure}

Taking the perpendicular velocity coordinate into account as well, we obtain a crescent plot for the final beam-ion distribution (\cref{figCrescentMed}). The phase space of background ions has stayed more Maxwellian in comparison because the dominant A/F waves are resonant with the background gyromotion. The centre of the beam crescent determines the optimal reference frame for a description in terms of pure pitch-angle scattering. Note that, because of acceleration of the background and the scattering waves, this optimal frame is neither the initial rest frame of the background ($v_x=0$) nor the one with the velocity at which A/F waves initially propagate ($v_x\sim 1\,v_A$); rather, it is approximately the centre-of-mass frame of the background medium after it has ceased to undergo acceleration at the end of saturation ($v_x\sim2.3\,v_A$), or alternatively the rest frame of the F/A waves. As a fast magnetosonic wave in the background medium, the F/A mode propagates at super-Alfv\'enic speed in high-$V_b$ scenarios.

\subsection{Run II: Low beam density}

\begin{figure}
\centering
\includegraphics[scale=.95]{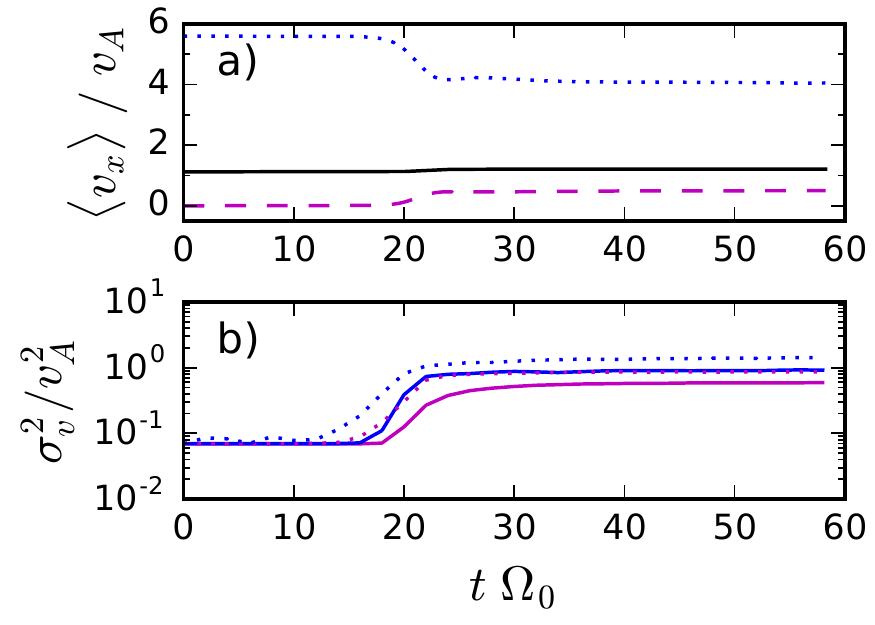}
\caption{Run II: Evolution of bulk $x$ velocities and velocity dispersion.}
\label{figEvolLow}
\end{figure}

With beam density and beam velocity reduced to $V_b=5.6\,v_A$ and $n_b/n_0=0.25$, the perpendicular temperature increase of the beam ions still begins at $t\,\Omega_0\approx10$, but is significantly slower (\cref{figEvolLow}a). The parallel velocities only begin to converge at $t\,\Omega_0\approx18$. At that point the F/A instability, similar to the A/F mode in the previously considered medium-density case, scatters background and beam ions such that the difference between the bulk velocities decreases while the combined momentum remains constant. When the saturation stage ends at $t\,\Omega_0\approx23$, the average background-ion velocity has gone up to $0.5\,v_A$; the average beam velocity has dropped to $4.0\,v_A$ in the simulation frame.

Using the cold-plasma dispersion relation, we would find that, with this velocity difference, the F/A mode should still be growing at a rate of $\gamma=0.4\,\Omega_0$. Taking the heating (\cref{figEvolLow}b) into account, a linear warm-beam calculation yields a growth rate of $\gamma=0.2\,\Omega_0$, which gets further suppressed in the simulation because of the nonlinear magnetic field.

\begin{figure}
\centering
\includegraphics[scale=.95]{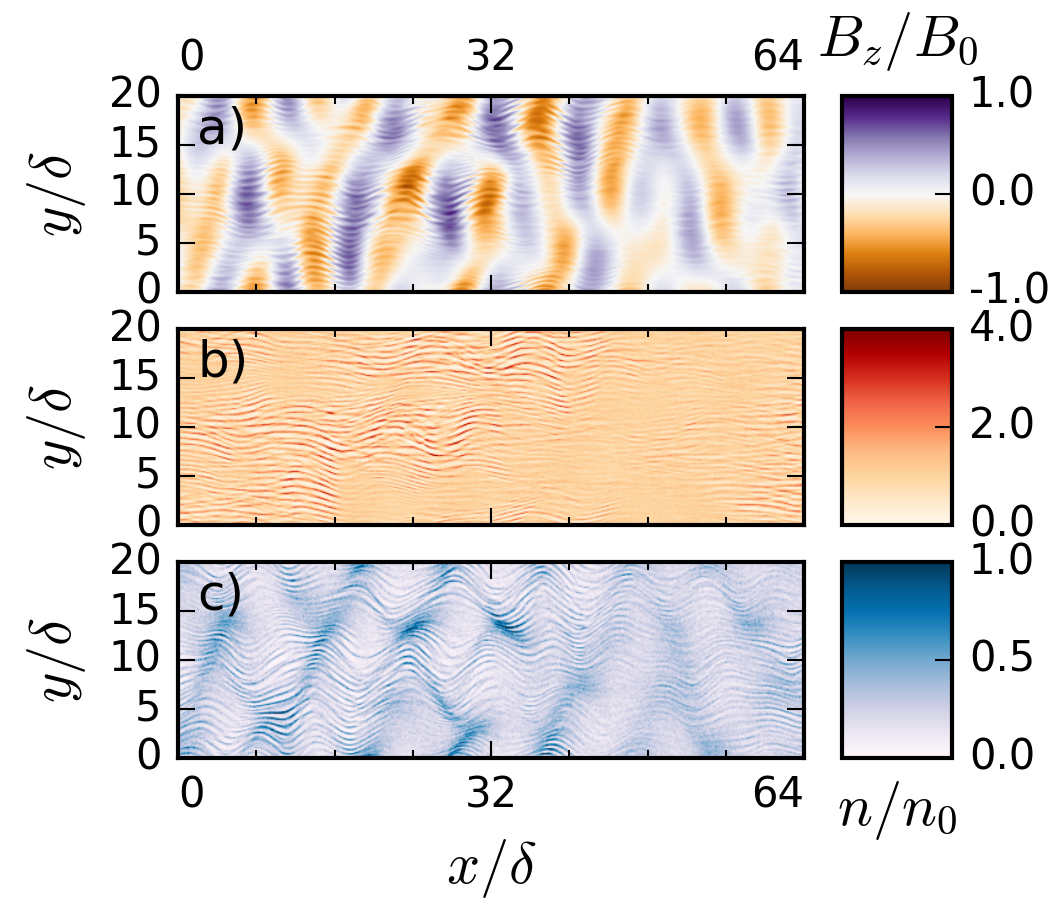}
\caption{Run II: Magnetic-field (a), background-density (b), and beam-density (c) profiles at $t\,\Omega_0=20$.}
\label{figProfileLow}
\end{figure}

At the onset of saturation, the profiles in \cref{figProfileLow} show that the magnetic field is dominated by the longitudinal F/A mode. The background-density profile, however, develops a transverse filamentation instability with $k_\perp\delta\approx6$ that is closely related to the interface instability investigated by \citet{winske90}. The beam density adopts a superposition of both the longitudinal and the transverse-filamentary structures.

Since the background species has a much larger charge density, its evolution during saturation influences the process more than in Run~I. In Run~II, the background ions are not accelerated by the left-handed A/F waves very effectively; that mode is relatively weak in this regime. The parallel velocity and temperature of the background thus change relatively slowly. Since the dominating F/A waves are right-hand polarised in the simulation frame ($\omega \approx+2.5\,\Omega_0$), they are far from resonance with the left-handed gyromotion of the background ions, but they do affect their trajectories nonetheless through perpendicular scattering. The direction of the motional electric field $\mathbf v \times \mathbf B$ varies wildly over one gyroperiod; hence the perpendicular velocity dispersion increases faster than along the magnetic field. Minute inhomogeneities in the charge density along the $y$ axis diffract the magnetic-wave planes by small angles. The perpendicular compression due to oblique F/A waves combined with gyoromotion around the local mean-field direction leads to the formation of density filaments, which wind around $\vec B_0$ helically. These filaments contract through magnetic pinching and also induce complementary filamentation in the counter-streaming beam medium.

\begin{figure}
\centering
\includegraphics[scale=.95]{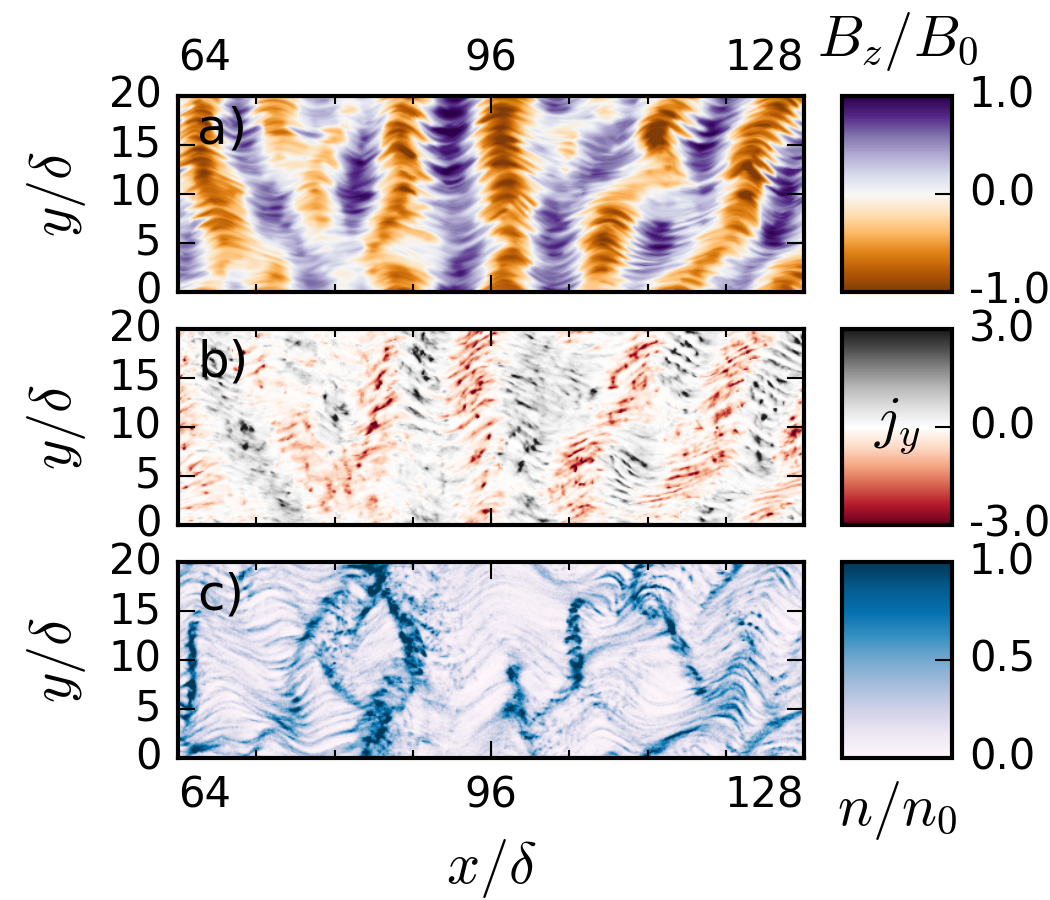}
\caption{Run II: Magnetic-field (a), background current-density in units of $n_0\,v_A$ (b), and beam density (c) at $t\,\Omega_0=23$.}
\label{figBgFilaments}
\end{figure}

As the magnetic field grows and ions get scattered perpendicularly, the helical filaments twist around the $x$ axis with an increasing radius. Eventually, the perpendicular fields grow so strong and the filaments' orientation becomes so oblique that self-focusing pinching forces form current filaments streaming up or down along the $y$ axis, a mechanism similar to the Weibel--Fried instability \citep{weibel59,fried59} or the mirror instability. In other words, fieldlines reconnect where the perpendicular currents grow strong enough; in three-dimensional geometry, this reconnection forms current rings or loops. These perpendicular currents, which still drift along $x$ with the background-ion velocity, interfere with the passing magnetic field of the dissipating F/A waves. Where this interference is constructive, the combined perpendicular magnetic field can become so strong as to trap streaming beam ions and create small transient shocks in the beam density (\cref{figBgFilaments}).

\begin{figure}
\centering
\includegraphics[scale=.95]{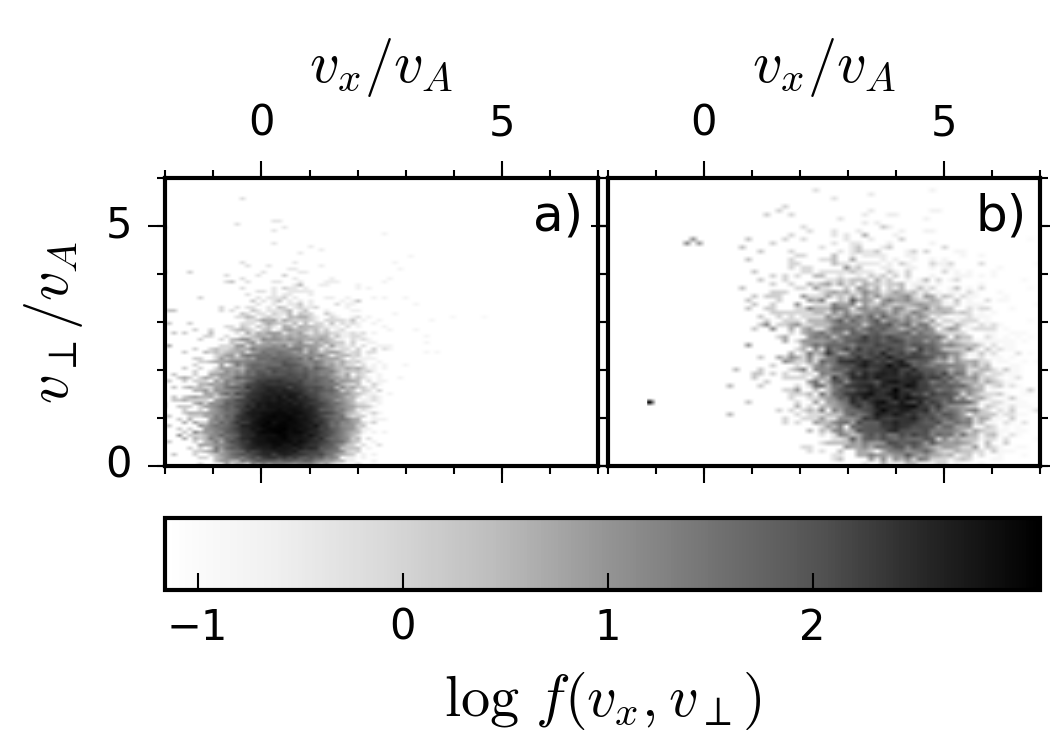}
\caption{Run II: Velocity phase space of a) background and b) beam ions at $t\,\Omega_0=30$.}
\label{figPhaseLow}
\end{figure}

One might expect that the formation of these `beam shocks' further reduces the beam velocity. On the contrary, global momentum exchange between the two ion species ends at this point (\cref{figEvolLow}), together with the growth of the magnetic field. The ion densities and the magnetic field have become too inhomogeneous to carry an Alfv\'en wave at the wavelength of the A/F or F/A type. Without gyroresonant waves to scatter them efficiently, the velocity distributions of both species stay approximately constant over the timescales relevant to this paper, with moderately anisotropic shapes as shown in \cref{figPhaseLow}.

The filamentation of both ion densities also occurs in Run~I, but not as strongly as in Run~II (see \cref{figProfileMed2}c). In the A/F dominated Run~I, the oblique F/A waves are too weak to efficiently scatter background ions perpendicularly and compress them, as evidenced by their isotropic velocity dispersion throughout saturation (\cref{figSatEvol}b). On the other hand, the beam ions are scattered by the strong A/F waves and do exhibit perpendicular compression early during saturation (\cref{figDProfile}b). The magnetic field of these dense beam-ion clumps in turn creates some filamentation in the background medium as well. In this sense, the left- and right-handed regimes mirror each other with the roles of background and beam reversed. However, the larger density of background ions means the regimes are not completely symmetric. In Run~I parallel acceleration decreases the velocity difference between both species very efficiently, and both F/A and A/F modes saturate before filamentation becomes as extreme as in Run~II.

\subsection{Run III: High beam density}

\begin{figure}
\centering
\includegraphics[scale=.95]{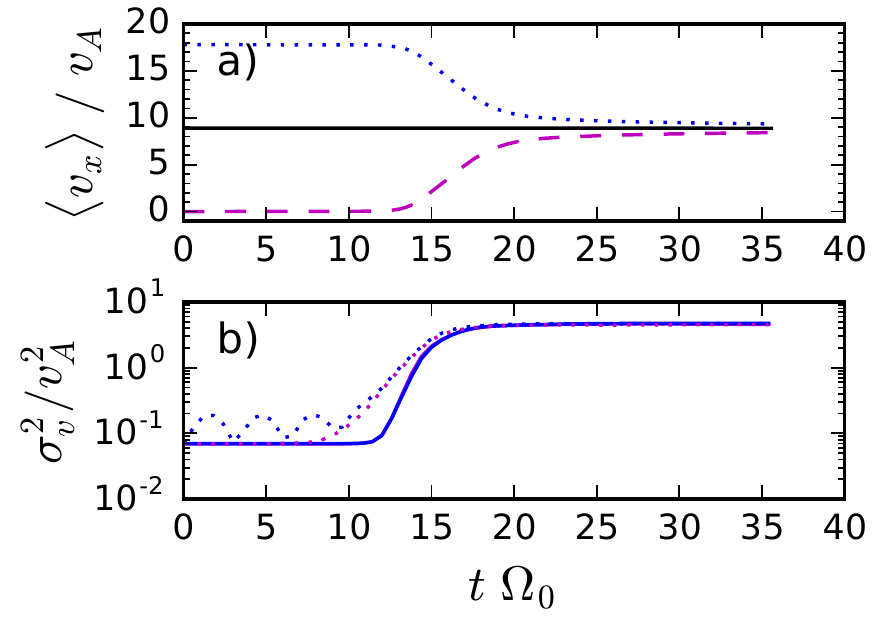}
\caption{Run III: Evolution of bulk $x$ velocities and velocity dispersions.}
\label{figEvolHi}
\end{figure}

In Run~III, the symmetry of left- and right-handed modes at the onset of saturation results in opposite but equal acceleration of background and beam ions (\cref{figEvolHi}a). Given our symmetric choice of ion densities $n_b = n_0$, of course, this is again equivalent to the conservation of total momentum. During the saturation stage, the initial beam velocity $V_b=17.8\,v_A$ drops to $9.4\,v_A$, while the bulk velocity of the background eventually reaches $8.4\,v_A$. In linear theory, this difference of $1.0\,v_A$ corresponds to the threshold for exciting the F/A or A/F instabilities. Although further turbulent mixing of both species occurs locally, it is no longer mediated by a global ion instability spanning the whole domain. Similarly, the heating rates for both species are the same during the saturation stage. Although perpendicular heating commences earlier, both temperature profiles are isotropic at the end of saturation.

\begin{figure}
\centering
\includegraphics[scale=.95]{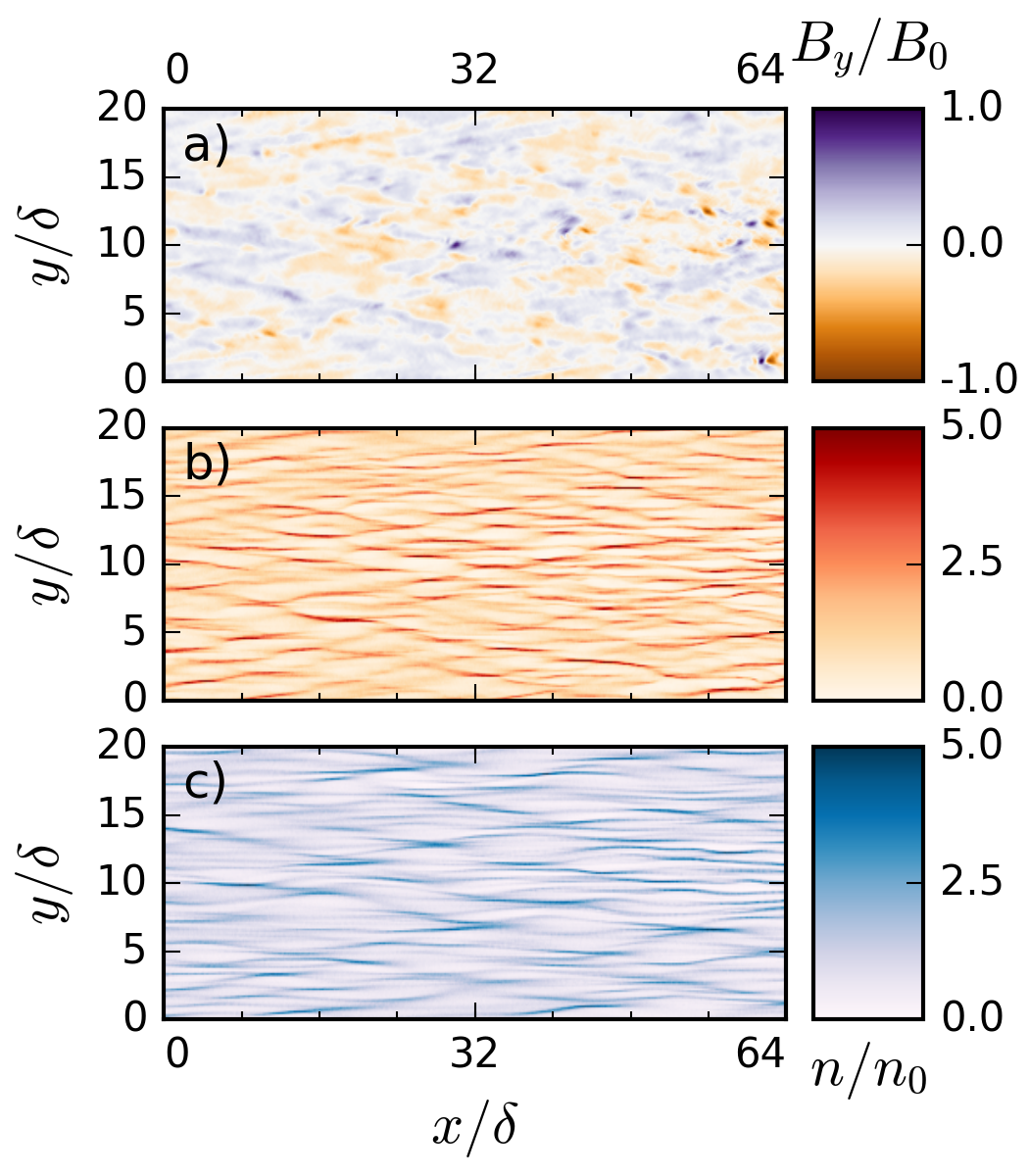}
\caption{Run III: Profiles at $t\,\Omega_0=12.0$ of a) $B_y$ component, b) background-ion density, c) beam-ion density}
\label{figProfileHi}
\end{figure}

A high degree of isotropy also characterises the magnetic turbulence very early during saturation (\cref{figProfileHi}a). Whereas the nonlinear stages of Runs~I and II were dominated by only one long-wavelength mode, viz.\ the A/F and the F/A mode, respectively, the spectrum of Run~III contains both modes with equal power. Consequently actual turbulence can develop between counter-streaming wave packets of nonlinear amplitude, resulting in a rapid growth of perpendicular modes. The well-defined longitudinal wave fronts in the perpendicular magnetic-field components disappear early, at only moderate field strengths.

From a kinetic point of view, the growth of perpendicular modes can be attributed to both ion species encountering right-hand polarised fluctuations in the magnetic field, which are strongly out of phase with their gyromotion --- the A/F waves for the beam, the F/A waves for the background. Hence both species are subject to the same perpendicular heating, compression, and filamentation that was visible more easily in the right-hand dominated regime. \Cref{figProfileHi} shows that the density profile of background and beam ions exhibits a filamentary structure similar to the one seen in \cref{figProfileLow}b for the F/A dominated Run~II. Simultaneously, like in the A/F dominated Run~I, the left-handed mode in each species' frame drives a parallel density modulation, which now shortens the length of the density filaments along the $x$ direction compared to the RHI-dominated case.

\begin{figure}
\centering
\includegraphics[scale=.95]{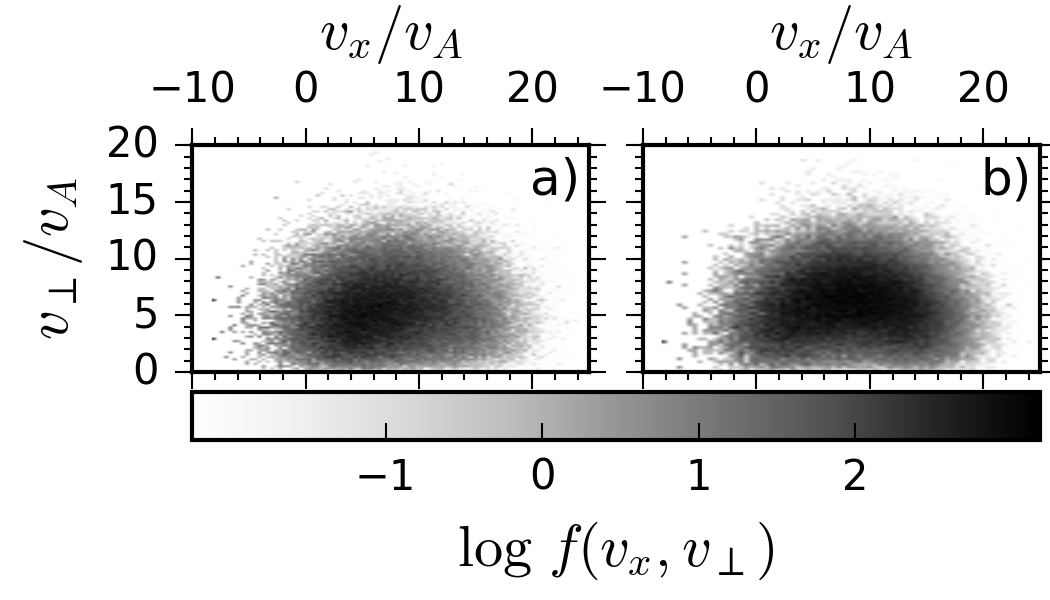}
\caption{Run III: Velocity-space density at $t\,\Omega_0=30.0$.}
\label{figCrescentHi}
\end{figure}

Since both perpendicular and parallel heating combine effectively, ions of both species are scattered rapidly by the isotropic magnetic turbulence, leading to true thermal equilibrium. As shown in \cref{figCrescentHi}, the two phase-space densities look almost identical at the end of Run~III. When viewed as one ion population, the final distribution is the result of a saturating firehose instability, since our symmetric mixed-turbulence setup is essentially indistinguishable from that.

\section{Discussion}
\label{secDiscussion}

\subsection{Effects of background gyroresonance}

The three saturation regimes for ion-beam instabilities differ by whether the higher-density background species is mainly accelerated by the dominant mode parallel to the mean field, scattered perpendicularly, or experiences both to a similar degree. Of course, what we in this article, with a perspective focused on the rest frame of the background, refer to as gyroresonant acceleration is from a more wave-centric point of view completely equivalent to magnetic trapping.

From a wave perspective, the regimes differ by whether the magnetic field during the linear-growth stage is more strongly gyroresonant with the background ions or the beam ions. This is completely determined by how much parallel acceleration each species experiences to decrease the relative drift below the threshold for ion-beam instabilities, while conserving the total momentum of the system. The stronger the accelerating force that acts on the background species is relative to the force on the beam, the more left-handed modes dominate over right-handed modes.

We have assumed a collisionless, perfectly quasineutral plasma with a zero-current initial state. Since both ion species can only couple to the magnetic field, and thus indirectly to each other, through their gyration, all growing electromagnetic modes are resonant with either gyrofrequency in the respective ions' frame of reference. Thus global field growth via any of these modes will inevitably saturate by rapid resonant acceleration as $\delta B \gtrsim 2\, B_0$.

If the initial state is not current-free but consists of a cosmic-ray beam and a background electron-ion plasma at rest, a two-stream instability will create electrostatic fluctuations, which drag the electrons along until the zero-current condition is fulfilled \citep{buneman58}. Since this compensating electron current will be established on time scales of the electron plasma frequency and length scales of the electron skin depth, however, the physics of this process are not captured by the standard MHD assumptions and hence by any MHD simulation. Within the framework of MHD, the single possible compensation mechanism for an external current $j_{\mathrm{CR}}$ is a helical magnetic field with the only characteristic length scale the system has, i.e.\ $\lambda\sim B_0/j_{\mathrm{CR}}$. Indeed, that prediction describes the filamentation process that we have observed during Run~II as well as a MHD model can, but it does not include the gyroresonant acceleration of the background that is responsible for the saturation of the left-handed regime. Nor does it contain the ion heating and diminishing growth of the short-wavelength mode.

The two-stream instability will not only drag the electrons, but also heat them in the process. \citet{zweibel10} have emphasised the importance of the electron temperature in their appendix, and although it is irrelevant for parallel-propagating modes (at least for $\omega\ll|\Omega_e|,\omega_p$ in either ion frame), it is possibly a crucial parameter for the filamentation in the nonlinear phase of the F/A instability. As a compressional phenomenon, the filaments couple to ion-acoustic oscillations and have been observed as `ion clumps' in 1D hybrid simulations \citep{terasawa88}. Since the electron pressure acts as a restoring force for these oscillations, increasing the electron temperature is likely to advance the formation of these filaments and thus possibly saturation of the right-handed regime.

\subsection{The left-hand resonant (Bell) instability}
\label{subBell}

The derivation of the instability described by \citet{bell04} neglects the background ions entirely. This instability can therefore not strictly be described as background-gyroresonant. Instead it assumes that the CR population in the upstream medium is hot enough (or follows a power-law distribution to large enough momentum) that some CR ions stream in the electron frame opposite to the bulk of the CRs, which we will assume to flow to the right. If the drift between any left-streaming CRs and the electrons exceeds $v_A$, these CRs will couple to the electrons in exactly the same way as the background ions do in our model and thus drive negative-helicity G/E waves via `left beam-tail resonance'. This scenario was previously discussed by \citet{gary84} as the left-hand resonant instability or LHI.

The expression `Bell instability' has been used in a wider sense by the astrophysical community to describe modes which are background-gyroresonant \citep[e.g. the cold-ion model in][see our remarks in the \cref{appendix}]{zweibel10}. In this sense, both our G/E and A/F modes are Bell instabilities. If the requirement for a mode to be `nonresonant' is taken as fulfilling $k_\res V_b/\Omega_b>1$ instead of not fulfilling condition~\eqref{eqnGyroresonance}, then our G/E mode becomes nonresonant if $m_b\, n_b\, V_b^2 > n_e\, v_A^2$, but this inequality says little about which physical process mediates the wave-particle coupling.

In Bell's original hot-beam configuration, a broad spectrum of waves with left-hand polarisation (in the background frame) is excited. It is possible that the saturation of this configuration towards equilibrium differs from our left-handed regime, an issue which further research should address. Since the background density is always larger than the density of left-streaming CRs, however, the background-gyroresonant modes will dominate the spectrum of left-hand polarised waves unless the relative drift between background ions and electrons is too small to excite the NRI. In the environment of supernova remnants, the NRI is definitely excited; the velocity of CRs certainly surpasses the Alfv\'en speed by more than a factor of $(n_e/n_0)^{1/2}$.

\subsection{Relation to parallel collisionless SNR shocks}

The distribution of SNR ejecta is neither as homogeneous nor as periodic as in our simulations, although the density of CR particles may be almost constant outside the free-escape boundary \citep{caprioli10}. Within that boundary, the average CR velocity rises and the CR density decreases with increasing distance from the SNR. Where the velocity is aligned with the interstellar magnetic field, parallel ion-beam instabilities will inevitably grow. After the A/F and F/A modes have reduced the relative drift between the CR beam and the ISM, denser SNR ejecta that have not yet been slowed down as much may catch up with and flow into a patch of saturated ISM plasma, driving it unstable again. This constant influx of CRs in an unsaturated state can amplify the magnetic field on ion scales beyond what we measure in our simulations, although presumably not by much since the velocity difference between both species (and thus the growth rate) is constantly decreasing.

The physics of parallel-shock formation at SNR, however, are easily understood from our results. We have shown that the F/A instability creates shocklets in the beam density when it grows nonlinear and saturates. Far into the left-handed regime, the structures created by the nonlinear A/F mode in the background-ion density are similarly shock-like, although their energy and density supply is too limited in our periodic simulations to develop into shocks in the solitonic, quasi-steady state described by the Rankine--Hugoniot conditions.

\citet{quest88} suggested that parallel collisionless shocks are created where conditions that allow the NRI to grow suddenly relax into a right-hand dominated parameter regime. Right-handed polarisation is commonly observed upstream of a shock, while \citet{krauss93} observed left-handed polarisation in the downstream medium. Taken together and viewed in the ISM frame, it follows that saturating A/F waves accelerate ISM protons on the downstream side. In the upstream plasma, saturating F/A waves decelerate CR protons, compressing the plasma which thus becomes unstable to the left-handed regime. As this regime saturates, the upstream medium turns into the downstream medium and the shock front passes by. In between, mixed turbulence creates the short filaments described by \citet{winske90} as the interface instability, depicted in our \cref{figProfileHi}.

The saturation of A/F waves in the left-handed regime results in a very efficient reflection of beam ions, as we saw in \cref{figCrescentMed}. Streaming further into a downstream plasma where A/F waves reached saturation amplitudes only recently, the reflected CRs can easily drive the plasma unstable again and are reflected a second time, gaining energy in the process (as measured in the original ISM rest frame). Thus the saturation mechanism we describe is closely connected to diffusive shock acceleration, particularly in the left-handed regime. Since the saturation stage lasts only a few inverse gyrofrequencies, however, pitch-angle scattering  at these short, non-diffusive timescales becomes important \citep{weidl15}, especially to explain how particles are injected into the DSA process.

If the plasma is sufficiently cold, the fast formation of the E/G and G/E modes contributes to an initial increase in the magnetic-field strength. This is likely to occur at the outermost SNR ejecta, where the surrounding ISM is still relatively quiescent and the CR beam will have a moderate velocity spread and uniform density. If the density of the CRs varies significantly on scales comparable to the long wavelengths of the F/A and A/F modes, these modes will not be able to form and coupling will only occur on short E/G and G/E length scales. But as we have shown, during saturation both species are heated to temperatures for which the whistler-type modes grow inefficiently. The long-wavelength A/F and F/A modes are therefore likely to be more important for DSA in strongly amplified fields than either whistler-type mode.

\section{Summary}

We have shown that a cold ion beam propagating in a cold magnetised background parallel to the magnetic field can, for $\omega\ll|\Omega_e|<\omega_p$, excite four instabilities: an electron-whistler wave excited by the gyromotion of the background ions (G/E), an electron-whistler excited by the gyromotion of the beam ions (E/G), a left-handed shear-Alfv\'en wave excited by the beam gyromotion (A/F), and a right-handed fast magnetosonic mode excited by the background gyromotion (F/A). Our notation stresses that these modes are mutually symmetric as one transforms the problem into the rest frame of the beam, in the sense that E/G becomes G/E and \emph{vice versa}, and A/F becomes F/A and \emph{vice versa}. If the beam velocity is only moderately super-Alfv\'enic, the two left-handed modes merge into one to form the NRI. For large beam drifts, the G/E mode is identical with the Bell instability.

Depending on the speed and density of the beam, it is possible that the A/F mode grows fastest until saturation (left-handed regime), that the F/A mode grows fastest (right-handed regime), or that both modes have comparable growth rates and interact with each other through a turbulent cascade while they are still growing (mixed-turbulence regime). In any case, the instabilities saturate by accelerating the two ion populations to reduce their relative drift below the marginal threshold. In the left-handed regime, the strong acceleration on the background population results from gyrobunched, relatively planar wave fronts in the background density. Saturation of the right-handed regime is preceded by the formation of self-focusing transverse filaments in both ion populations, which often develop into current loops and even small quasi-perpendicular shocklets in the beam density. The mixed-turbulence case resembles the firehose instability and leads to the appearance of short filaments often observed at the interfaces of parallel and quasi-parallel shocks.

Since the ion kinetics of parallel beam instabilities are already so rich in variety, this article does not address the physics on electron scales. We expect the filamentation in the right-handed regime to be particularly sensitive to the electron temperature. In the left-handed regime, hotter electrons may lead to the emergence of kinetic Alfv\'en waves \citep{malovichko15} and amplify ion-density filamentation as well. Whether and how this affects the scattering of beam ions during saturation is extremely important to determine the rate of downstream cosmic-ray reflection in diffusive shock acceleration. The framework we present is likely also applicable to simulations of galactic feedback of CRs in the ISM \citep{ruszkowski17,wiener17,mao18}.

The filamentation behaviour and the interaction of oblique modes may also differ in subtle ways in a more realistic 3D geometry. In current laboratory experiments using super-Alfv\'enic carbon-ion beams from laser-target interactions, the F/A and E/G instabilities have already been measured \citep{heuer18}. Future work will focus on identifying the ion-density filamentation that our simulations predict in this regime, as well as accessing the mixed-turbulence and even the left-handed regime experimentally.

\section*{Appendix: The background-{\allowbreak}gyroresonant modes}
\label[appendix]{appendix}

\newcommand{\EEpert}{\EE'}
\newcommand{\BBpert}{\BB'}

Our goal is to show that the so-called nonresonant instability is actually due to an \emph{almost} resonant interaction between the gyrofrequency of the background ions and an Alfv\'en wave in the plasma, in other words, the NRI is an ion-cyclotron wave when viewed in the background frame. In the rest frame of the beam or cosmic-ray ions, the blue-shifted synchrotron radiation of the leftwards-streaming background ions excites short-wavelength whistler waves.

Starting from the momentum equations for cold background and beam ions ($\alpha=0,b$) and electrons ($\alpha=e$), and the wave equation for a current-driven EM wave,
\begin{align*}
\mathrm D_{\alpha} \vec v_\alpha &= \Omega_\alpha\ (\EE + \vec v_\alpha/c \times \BB)\ B_0^{-1},\\
\vec j_\alpha &= q_\alpha n_\alpha \vec v_\alpha,\\
\nabla \times (\nabla \times \EE) &= -c^{-2}\, \left(\partial_t^2\, \EE + c\, \partial_t \sum_\alpha \vec j_\alpha\right),
\end{align*}
one can quickly derive the dispersion relation of parallel Alfv\'enic waves for charged-particle beams with a relative drift $V_\alpha$ parallel to the magnetic field \citep[e.g.][]{akhiezer1975}. The convective derivative is defined as $\mathrm D_{\alpha} = \partial_t + V_\alpha \partial_z$. Performing a Fourier transform with argument $k_\parallel$ in $z$ space and a Laplace transform with argument $\II \omega$ in time (and taking $\lim \vec v_\alpha(0^+) \to V_\alpha\, \hat z$), we find for the transformed perpendicular field components $\vec E_\perp = (\tilde E_x, \tilde E_y)^\top$
\begin{multline*}
\left(\omega^2 - c^2\,k_\parallel^2\right)\, \vec E_\perp =\\ \sum_\alpha \omega_{p,\alpha}^2\ \frac{\omega-V_\alpha k_\parallel}{(\omega-V_\alpha k_\parallel)^2-\Omega_\alpha^2} \left( \begin{matrix} \omega-V_\alpha k_\parallel & -\II \Omega_\alpha \\ \II \Omega_\alpha & \omega-V_\alpha k_\parallel \end{matrix} \right) \vec E_\perp.
\end{multline*}
Diagonalising this equation by introducing $E_{\mathrm{R/L}}=\tilde E_x\pm\II \tilde E_y$, one obtains
\begin{equation*}
\left(\omega^2 - c^2\,k_\parallel^2\right)\, E_{\mathrm{R/L}} = \sum_\alpha \omega_{p,\alpha}^2 \frac{\omega-V_\alpha k_\parallel}{\omega-V_\alpha k_\parallel \pm \Omega_\alpha} E_{\mathrm{R/L}}.
\end{equation*}
Our interpretation of negative frequency and wavenumber allows us to drop the distinction between the two signs: taking the negative sign is equivalent to inverting $\omega$ and $k_\parallel$ in the equation for $E_\mathrm{R}$. In the relativistic regime, $\Omega_\alpha$ is simply substituted by the relativistic gyrofrequency $\Omega_\alpha\,(1-V_\alpha^2/c^2)^{1/2}$. The standard hybrid-model assumptions described in \cref{secTheory} then yield equation~\eqref{eqnBeamDR}.

For the `nonresonant' instability, $|\omega|^2 \ll \omega_p^2$ holds so that in the electron rest frame
\begin{equation*}
\left( \frac{\omega}{\Omega_0} - k_\parallel^2 \delta^2 \right) n_e \approx n_0 \frac{\omega-v_0 k_\parallel}{\omega-v_0 k_\parallel +\Omega_0} + n_b \frac{q_b^2}{m_b}\, \frac{\omega-v_b k_\parallel}{\omega-v_b k_\parallel + \Omega_b}.
\end{equation*}
Since we are not looking for beam-gyroresonant modes, we can drop the last term because its denominator is large and $n_b$ is small. To maximise the growth rate $\gamma=\Im\omega$, we look at the imaginary part of the remaining equation:
\begin{equation}
\frac{\gamma}{\Omega_0} \left[ (\varpi-v_0 k_\parallel + \Omega_0)^2 + \gamma^2 \right] = \frac{n_0}{n_e}\, \gamma\, \Omega_0,
\tag{$\star$}
\label{eqnImDR}
\end{equation}
where $\varpi=\Re\omega$. It is now patently obvious that $\gamma^2$ reaches its maximum $\overline\gamma^2$ in the case of (exact) background gyroresonance,
\begin{equation*}
\varpi - v_0 k_\parallel + \Omega_0 = 0,
\end{equation*}
whence it follows immediately that $\overline \gamma = (n_0/n_e)^{1/2}\, \Omega_0$. Inserting this growth rate and the background-gyroresonance condition into the real part of the equation yields
\begin{equation*}
k_\parallel = \frac12 \frac{v_0}{\Omega_0 \delta^2} \left( 1 \pm \sqrt{1-\left(1+\sqrt\frac{n_0}{n_e}\right)
\frac{4\,\Omega_0^2\,\delta^2}{v_0^2}}\right).
\end{equation*}

For non-evanescent propagation of left-handed modes with growth rate $\overline \gamma$, the drift between background ions and electrons must be large enough that the discriminant of this equation is positive. Converted to the beam velocity $V_b$ as measured in the background frame, this implies the following condition on the beam current:
\begin{equation*}
e\,q_b\,n_b\,V_b > j_{\mathrm{crit}} = 2\, \sqrt{1+ \sqrt\frac{n_0}{n_e} }\, e\, n_e\, v_A \approx 2.8\, e\, n_0\, v_A.
\end{equation*}

As long as this condition is fulfilled, we can find two solutions with $\gamma=\overline\gamma$ and $k_\parallel\in\mathbb R$, viz.\ a long-wavelength solution corresponding to our A/F mode at $k_\parallel\approx2\,\Omega_0/v_0$ and a short-wavelength solution:
\begin{equation*}
k_\res = \frac{v_0}{\Omega_0\, \delta^2} + \mathcal O \left(\frac{v_A}{v_0}\right).
\end{equation*}
In the low-beam-density limit, electron and background-ion density are almost identical, and current neutrality implies that $n_0 v_0 = -q_b n_b v_b$. Hence we can use $\omega_p^2/\Omega_0 \approx n_0 e/B_0$ to write (for $q_b=1$)
\begin{equation*}
k_\res \approx \frac{v_0\, n_0\, e}{c^2\, B_0} = - \frac{e\, n_b\, v_b}{B_0},
\end{equation*}
which is exactly our E/G mode and almost equivalent to the fastest-growing mode of the `nonresonant' instability described by \citet{winske84} and \citet{bell04} ---  up to a factor of \textonehalf.

This `missing' factor is due to our postulation that the beam current is greater than $j_\mathrm{crit}$; we posited that there exist spatially periodic solutions with the maximal growth rate $\overline\gamma$. At lower current densities, left-handed unstable modes may still exist but grow at a slower rate. If $0<\gamma<\overline\gamma$, equation \eqref{eqnImDR} shows that $\varpi-v_0k_\parallel\neq-\Omega_0$. One can then make the \emph{a priori} assumption $|\omega-v_0 k_\parallel| \ll \Omega_0$ \citep[e.g.][]{zweibel10} and see if there are any modes that look almost purely growing when Doppler-shifted to the background frame. In that case, one expands the background-gyroresonant term to second order in $\omega' = \omega-v_0 k_\parallel$ and ends up with
\begin{equation*}
\omega' = - \frac{q_b n_b}{2\, n_0} \Omega_0 \pm \frac{\Omega_0\, n_e}{2\, n_0} \sqrt{\left(\frac{q_b n_b}{n_e}\right)^2 - 4 \frac{n_0}{n_e} \frac{v_0\,k_\parallel}{\Omega} + 4 \frac{n_0}{n_e} \delta^2 k_\parallel^2}.
\end{equation*}

This approach yields only one fastest-growing mode at the extremum of the discriminant, $k_\mathrm{Bell}=k_\res/2$, the real part of our solution for $k_\parallel$ if the beam current is sub-critical. The growth rate then becomes
\begin{equation*}
\gamma_\mathrm{Bell} = \frac{\Omega_0}{2} \frac{q_b\,n_b}{n_0} \sqrt{\frac{n_e}{n_0}\left(\frac{v_b}{v_A}\right)^2 -1} \longrightarrow \frac{\Omega_0}{2} \frac{q_b\,n_b}{n_0} \frac{V_b}{v_A}
\end{equation*}
in the limit of low beam density and large beam velocity. But this solution is only compatible with the \emph{a priori} assumption that made the Taylor expansion possible if the beam current is weaker than $j_\mathrm{crit}$. If the current is stronger, one must limit the space of solutions to the ones compatible with $|\omega-v_0k_\parallel|<\Omega_0$. Instead of one broad unstable $k_\parallel$ range centred around the extremum of the discriminant, one then obtains two narrow bands at equal distance from $k_\mathrm{Bell}$ --- our G/E and A/F modes.

The advantage of this approach is that it shows that left-handed modes become unstable even before exact gyroresonance is possible, i.e.\ if the beam current is below $j_{\mathrm{crit}}$ but strong enough for $v_b>v_A(n_0/n_e)^{1/2}$. As long as the real part of the dispersion relation allows for a solution $(\omega,k_\parallel)$ with $|\varpi-v_0k_\parallel+\Omega_0|<(n_0/n_e)^{1/2}\,\Omega_0$, its imaginary part \eqref{eqnImDR} yields a strictly positive growth rate $\gamma$. The expansion in $|\omega-v_0k_\parallel|/\Omega_0$ is therefore an appropriate approximation in the medium-velocity regime in which the G/E and A/F modes merge into one peak, which we called the NRI in Run~I. This `nonresonant' instability thus couples the gyromotion of background ions to plasma waves even if the gyroresonance condition can be fulfilled only approximately --- within a `margin of error' of about $\Omega_0$. However, this approach obscures the gyroresonant origin of the mode(s).

The beam-gyroresonant instabilities with positive helicity can be derived analogously if one drops the background-gyroresonant term from the dispersion relation.

\section*{Acknowledgments}
M.S.W. is supported by the Deutsche Forschungsgemeinschaft and thanks George~Morales for extremely valuable discussions during the gestation of this article and Damiano~Caprioli and Anatoly~Spitkovsky for useful comments. This work was partially supported by the Max-Planck/Princeton Center for Plasma Physics. Computations were performed on the Hoffman2 cluster at UCLA and made possible by the Bhaumik~Center for Theoretical Physics. This work was supported by the DTRA under Contract No. HDTRA1-12-1-0024, and the DOE under Contract No. DE-SC0017900.

\bibliographystyle{mnras}
\bibliography{BeamRefs}

\begin{thebibliography}{}
\makeatletter
\relax
\def\mn@urlcharsother{\let\do\@makeother \do\$\do\&\do\#\do\^\do\_\do\%\do\~}
\def\mn@doi{\begingroup\mn@urlcharsother \@ifnextchar [ {\mn@doi@}
  {\mn@doi@[]}}
\def\mn@doi@[#1]#2{\def\@tempa{#1}\ifx\@tempa\@empty \href
  {http://dx.doi.org/#2} {doi:#2}\else \href {http://dx.doi.org/#2} {#1}\fi
  \endgroup}
\def\mn@eprint#1#2{\mn@eprint@#1:#2::\@nil}
\def\mn@eprint@arXiv#1{\href {http://arxiv.org/abs/#1} {{\tt arXiv:#1}}}
\def\mn@eprint@dblp#1{\href {http://dblp.uni-trier.de/rec/bibtex/#1.xml}
  {dblp:#1}}
\def\mn@eprint@#1:#2:#3:#4\@nil{\def\@tempa {#1}\def\@tempb {#2}\def\@tempc
  {#3}\ifx \@tempc \@empty \let \@tempc \@tempb \let \@tempb \@tempa \fi \ifx
  \@tempb \@empty \def\@tempb {arXiv}\fi \@ifundefined
  {mn@eprint@\@tempb}{\@tempb:\@tempc}{\expandafter \expandafter \csname
  mn@eprint@\@tempb\endcsname \expandafter{\@tempc}}}

\bibitem[\protect\citeauthoryear{{Achterberg}}{{Achterberg}}{1983}]{achterberg83}
{Achterberg} A.,  1983, \aap, \href
  {http://adsabs.harvard.edu/abs/1983A%26A...119..274A} {119, 274}

\bibitem[\protect\citeauthoryear{{Akhiezer}}{{Akhiezer}}{1975}]{akhiezer1975}
{Akhiezer} A.~I.,  1975, {Plasma electrodynamics - Vol.1: Linear theory}.
{Pergamon Press}, {Oxford}

\bibitem[\protect\citeauthoryear{{Akimoto}, {Gary}  \& {Omidi}}{{Akimoto}
  et~al.}{1987}]{akimoto87}
{Akimoto} K.,  {Gary} S.~P.,   {Omidi} N.,  1987, \mn@doi [\jgr]
  {10.1029/JA092iA10p11209}, \href
  {http://adsabs.harvard.edu/abs/1987JGR....9211209A} {92, 11209}

\bibitem[\protect\citeauthoryear{{Akimoto}, {Winske}, {Gary}  \&
  {Thomsen}}{{Akimoto} et~al.}{1993}]{akimoto93}
{Akimoto} K.,  {Winske} D.,  {Gary} S.~P.,   {Thomsen} M.~F.,  1993, \mn@doi
  [\jgr] {10.1029/92JA02345}, \href
  {http://adsabs.harvard.edu/abs/1993JGR....98.1419A} {98, 1419}

\bibitem[\protect\citeauthoryear{{Amato} \& {Blasi}}{{Amato} \&
  {Blasi}}{2009}]{amato09}
{Amato} E.,  {Blasi} P.,  2009, \mn@doi [\mnras]
  {10.1111/j.1365-2966.2008.14200.x}, \href
  {http://adsabs.harvard.edu/abs/2009MNRAS.392.1591A} {392, 1591}

\bibitem[\protect\citeauthoryear{{Axford}, {Leer}  \& {Skadron}}{{Axford}
  et~al.}{1977}]{axford77}
{Axford} W.~I.,  {Leer} E.,   {Skadron} G.,  1977, International Cosmic Ray
  Conference, \href {http://adsabs.harvard.edu/abs/1977ICRC...11..132A} {11,
  132}

\bibitem[\protect\citeauthoryear{{Bell}}{{Bell}}{1978}]{bell78}
{Bell} A.~R.,  1978, \mn@doi [\mnras] {10.1093/mnras/182.2.147}, \href
  {http://adsabs.harvard.edu/abs/1978MNRAS.182..147B} {182, 147}

\bibitem[\protect\citeauthoryear{{Bell}}{{Bell}}{2004}]{bell04}
{Bell} A.~R.,  2004, \mn@doi [\mnras] {10.1111/j.1365-2966.2004.08097.x}, \href
  {http://adsabs.harvard.edu/abs/2004MNRAS.353..550B} {353, 550}

\bibitem[\protect\citeauthoryear{{Blandford} \& {Ostriker}}{{Blandford} \&
  {Ostriker}}{1978}]{blandford78}
{Blandford} R.~D.,  {Ostriker} J.~P.,  1978, \mn@doi [\apjl] {10.1086/182658},
  \href {http://adsabs.harvard.edu/abs/1978ApJ...221L..29B} {221, L29}

\bibitem[\protect\citeauthoryear{{Buneman}}{{Buneman}}{1958}]{buneman58}
{Buneman} O.,  1958, \mn@doi [Physical Review Letters]
  {10.1103/PhysRevLett.1.8}, \href
  {http://adsabs.harvard.edu/abs/1958PhRvL...1....8B} {1, 8}

\bibitem[\protect\citeauthoryear{{Caprioli} \& {Spitkovsky}}{{Caprioli} \&
  {Spitkovsky}}{2013}]{caprioli13}
{Caprioli} D.,  {Spitkovsky} A.,  2013, \mn@doi [\apjl]
  {10.1088/2041-8205/765/1/L20}, \href
  {http://adsabs.harvard.edu/abs/2013ApJ...765L..20C} {765, L20}

\bibitem[\protect\citeauthoryear{{Caprioli} \& {Spitkovsky}}{{Caprioli} \&
  {Spitkovsky}}{2014}]{caprioli14}
{Caprioli} D.,  {Spitkovsky} A.,  2014, \mn@doi [\apj]
  {10.1088/0004-637X/783/2/91}, \href
  {http://adsabs.harvard.edu/abs/2014ApJ...783...91C} {783, 91}

\bibitem[\protect\citeauthoryear{{Caprioli}, {Amato}  \& {Blasi}}{{Caprioli}
  et~al.}{2010}]{caprioli10}
{Caprioli} D.,  {Amato} E.,   {Blasi} P.,  2010, \mn@doi [Astroparticle
  Physics] {10.1016/j.astropartphys.2010.03.001}, \href
  {http://adsabs.harvard.edu/abs/2010APh....33..307C} {33, 307}

\bibitem[\protect\citeauthoryear{{Cipolla}, {Silevitch}  \& {Golden}}{{Cipolla}
  et~al.}{1977}]{cipolla77}
{Cipolla} Jr. J.~W.,  {Silevitch} M.~B.,   {Golden} K.~I.,  1977, \mn@doi
  [Physics of Fluids] {10.1063/1.861865}, \href
  {http://adsabs.harvard.edu/abs/1977PhFl...20..282C} {20, 282}

\bibitem[\protect\citeauthoryear{{Clark}, {Winske}, {Schaeffer}, {Everson},
  {Bondarenko}, {Constantin}  \& {Niemann}}{{Clark} et~al.}{2013}]{clark13}
{Clark} S.~E.,  {Winske} D.,  {Schaeffer} D.~B.,  {Everson} E.~T.,
  {Bondarenko} A.~S.,  {Constantin} C.~G.,   {Niemann} C.,  2013, \mn@doi
  [Physics of Plasmas] {10.1063/1.4819251}, \href
  {http://adsabs.harvard.edu/abs/2013PhPl...20h2129C} {20, 082129}

\bibitem[\protect\citeauthoryear{{Fairfield}}{{Fairfield}}{1969}]{fairfield69}
{Fairfield} D.~H.,  1969, \mn@doi [\jgr] {10.1029/JA074i014p03541}, \href
  {http://adsabs.harvard.edu/abs/1969JGR....74.3541F} {74, 3541}

\bibitem[\protect\citeauthoryear{{Fried}}{{Fried}}{1959}]{fried59}
{Fried} B.~D.,  1959, \mn@doi [Physics of Fluids] {10.1063/1.1705933}, \href
  {http://adsabs.harvard.edu/abs/1959PhFl....2..337F} {2, 337}

\bibitem[\protect\citeauthoryear{{Gargat{\'e}} \& {Spitkovsky}}{{Gargat{\'e}}
  \& {Spitkovsky}}{2012}]{gargate12}
{Gargat{\'e}} L.,  {Spitkovsky} A.,  2012, \mn@doi [\apj]
  {10.1088/0004-637X/744/1/67}, \href
  {http://adsabs.harvard.edu/abs/2012ApJ...744...67G} {744, 67}

\bibitem[\protect\citeauthoryear{{Gargat{\'e}}, {Fonseca}, {Niemiec}, {Pohl},
  {Bingham}  \& {Silva}}{{Gargat{\'e}} et~al.}{2010}]{gargate10}
{Gargat{\'e}} L.,  {Fonseca} R.~A.,  {Niemiec} J.,  {Pohl} M.,  {Bingham} R.,
  {Silva} L.~O.,  2010, \mn@doi [\apjl] {10.1088/2041-8205/711/2/L127}, \href
  {http://adsabs.harvard.edu/abs/2010ApJ...711L.127G} {711, L127}

\bibitem[\protect\citeauthoryear{{Gary}}{{Gary}}{1978}]{gary78}
{Gary} S.~P.,  1978, Nuclear Fusion, \href
  {http://adsabs.harvard.edu/abs/1978NucFu..18..327G} {18, 327}

\bibitem[\protect\citeauthoryear{{Gary}}{{Gary}}{1991}]{gary91}
{Gary} S.~P.,  1991, \mn@doi [\ssr] {10.1007/BF00196632}, \href
  {http://adsabs.harvard.edu/abs/1991SSRv...56..373G} {56, 373}

\bibitem[\protect\citeauthoryear{{Gary}, {Foosland}, {Smith}, {Lee}  \&
  {Goldstein}}{{Gary} et~al.}{1984}]{gary84}
{Gary} S.~P.,  {Foosland} D.~W.,  {Smith} C.~W.,  {Lee} M.~A.,   {Goldstein}
  M.~L.,  1984, \mn@doi [Physics of Fluids] {10.1063/1.864797}, \href
  {http://adsabs.harvard.edu/abs/1984PhFl...27.1852G} {27, 1852}

\bibitem[\protect\citeauthoryear{{Gleaves}, {Southwood}, {Dunlop}  \&
  {Mier-Jedrzejowicz}}{{Gleaves} et~al.}{1988}]{gleaves88}
{Gleaves} D.~G.,  {Southwood} D.~J.,  {Dunlop} M.~W.,   {Mier-Jedrzejowicz}
  W.~A.~C.,  1988, \mn@doi [Advances in Space Research]
  {10.1016/0273-1177(88)90130-5}, \href
  {http://adsabs.harvard.edu/abs/1988AdSpR...8..181G} {8, 181}

\bibitem[\protect\citeauthoryear{{Heuer} et~al.,}{{Heuer}
  et~al.}{2018}]{heuer18}
{Heuer} P.~V.,  et~al., 2018, \mn@doi [Physics of Plasmas] {10.1063/1.5017637},
  \href {http://adsabs.harvard.edu/abs/2018PhPl...25c2104H} {25, 032104}

\bibitem[\protect\citeauthoryear{{Huntington} et~al.,}{{Huntington}
  et~al.}{2015}]{huntington15}
{Huntington} C.~M.,  et~al., 2015, \mn@doi [Nature Physics]
  {10.1038/nphys3178}, \href
  {http://adsabs.harvard.edu/abs/2015NatPh..11..173H} {11, 173}

\bibitem[\protect\citeauthoryear{{Jian} et~al.,}{{Jian} et~al.}{2014}]{jian14}
{Jian} L.~K.,  et~al., 2014, \mn@doi [\apj] {10.1088/0004-637X/786/2/123},
  \href {http://adsabs.harvard.edu/abs/2014ApJ...786..123J} {786, 123}

\bibitem[\protect\citeauthoryear{{Krall} \& {Trivelpiece}}{{Krall} \&
  {Trivelpiece}}{1973}]{krall1973}
{Krall} N.~A.,  {Trivelpiece} A.~W.,  1973, {Principles of plasma physics}.
{McGraw-Hill}, {New York}

\bibitem[\protect\citeauthoryear{{Krauss-Varban} \& {Omidi}}{{Krauss-Varban} \&
  {Omidi}}{1993}]{krauss93}
{Krauss-Varban} D.,  {Omidi} N.,  1993, \mn@doi [\grl] {10.1029/93GL01125},
  \href {http://adsabs.harvard.edu/abs/1993GeoRL..20.1007K} {20, 1007}

\bibitem[\protect\citeauthoryear{{Kulsrud}}{{Kulsrud}}{2005}]{kulsrud2005}
{Kulsrud} R.~M.,  2005, {Plasma physics for astrophysics}.
{Princeton University Press}, {Princeton}

\bibitem[\protect\citeauthoryear{{Kulsrud} \& {Pearce}}{{Kulsrud} \&
  {Pearce}}{1969}]{kulsrud69}
{Kulsrud} R.,  {Pearce} W.~P.,  1969, \mn@doi [\apj] {10.1086/149981}, \href
  {http://adsabs.harvard.edu/abs/1969ApJ...156..445K} {156, 445}

\bibitem[\protect\citeauthoryear{{Lucek} \& {Bell}}{{Lucek} \&
  {Bell}}{2000}]{lucek00}
{Lucek} S.~G.,  {Bell} A.~R.,  2000, \mn@doi [\mnras]
  {10.1046/j.1365-8711.2000.03363.x}, \href
  {http://adsabs.harvard.edu/abs/2000MNRAS.314...65L} {314, 65}

\bibitem[\protect\citeauthoryear{{Lucek}, {Horbury}, {Dandouras}  \&
  {R{\`e}me}}{{Lucek} et~al.}{2008}]{lucek08}
{Lucek} E.~A.,  {Horbury} T.~S.,  {Dandouras} I.,   {R{\`e}me} H.,  2008,
  \mn@doi [Journal of Geophysical Research (Space Physics)]
  {10.1029/2007JA012756}, \href
  {http://adsabs.harvard.edu/abs/2008JGRA..113.7S02L} {113, A07S02}

\bibitem[\protect\citeauthoryear{{Malovichko}, {Voitenko}  \& {De
  Keyser}}{{Malovichko} et~al.}{2015}]{malovichko15}
{Malovichko} P.,  {Voitenko} Y.,   {De Keyser} J.,  2015, \mn@doi [\mnras]
  {10.1093/mnras/stv1533}, \href
  {http://adsabs.harvard.edu/abs/2015MNRAS.452.4236M} {452, 4236}

\bibitem[\protect\citeauthoryear{{Mao} \& {Ostriker}}{{Mao} \&
  {Ostriker}}{2018}]{mao18}
{Mao} S.~A.,  {Ostriker} E.~C.,  2018, \mn@doi [\apj]
  {10.3847/1538-4357/aaa88e}, \href
  {http://adsabs.harvard.edu/abs/2018ApJ...854...89M} {854, 89}

\bibitem[\protect\citeauthoryear{{Niemiec}, {Pohl}, {Stroman}  \&
  {Nishikawa}}{{Niemiec} et~al.}{2008}]{niemiec08}
{Niemiec} J.,  {Pohl} M.,  {Stroman} T.,   {Nishikawa} K.-I.,  2008, \mn@doi
  [\apj] {10.1086/590054}, \href
  {http://adsabs.harvard.edu/abs/2008ApJ...684.1174N} {684, 1174}

\bibitem[\protect\citeauthoryear{{Pelletier}, {Lemoine}  \&
  {Marcowith}}{{Pelletier} et~al.}{2006}]{pelletier06}
{Pelletier} G.,  {Lemoine} M.,   {Marcowith} A.,  2006, \mn@doi [\aap]
  {10.1051/0004-6361:20054737}, \href
  {http://adsabs.harvard.edu/abs/2006A%26A...453..181P} {453, 181}

\bibitem[\protect\citeauthoryear{{Quest}}{{Quest}}{1988}]{quest88}
{Quest} K.~B.,  1988, \mn@doi [\jgr] {10.1029/JA093iA09p09649}, \href
  {http://adsabs.harvard.edu/abs/1988JGR....93.9649Q} {93, 9649}

\bibitem[\protect\citeauthoryear{{Riquelme} \& {Spitkovsky}}{{Riquelme} \&
  {Spitkovsky}}{2009}]{riquelme09}
{Riquelme} M.~A.,  {Spitkovsky} A.,  2009, \mn@doi [\apj]
  {10.1088/0004-637X/694/1/626}, \href
  {http://adsabs.harvard.edu/abs/2009ApJ...694..626R} {694, 626}

\bibitem[\protect\citeauthoryear{{Roennmark}}{{Roennmark}}{1982}]{roennmark1982}
{Roennmark} K.,  1982, Technical report, {Waves in homogeneous, anisotropic
  multicomponent plasmas (WHAMP)}

\bibitem[\protect\citeauthoryear{{Russell} \& {Greenstadt}}{{Russell} \&
  {Greenstadt}}{1979}]{russell79}
{Russell} C.~T.,  {Greenstadt} E.~W.,  1979, \mn@doi [\ssr]
  {10.1007/BF00174109}, \href
  {http://adsabs.harvard.edu/abs/1979SSRv...23....3R} {23, 3}

\bibitem[\protect\citeauthoryear{{Ruszkowski}, {Yang}  \&
  {Zweibel}}{{Ruszkowski} et~al.}{2017}]{ruszkowski17}
{Ruszkowski} M.,  {Yang} H.-Y.~K.,   {Zweibel} E.,  2017, \mn@doi [\apj]
  {10.3847/1538-4357/834/2/208}, \href
  {http://adsabs.harvard.edu/abs/2017ApJ...834..208R} {834, 208}

\bibitem[\protect\citeauthoryear{{Sauer}, {Dubinin}, {Dunlop}, {Baumgartel}  \&
  {Tarasov}}{{Sauer} et~al.}{1999}]{sauer99}
{Sauer} K.,  {Dubinin} E.,  {Dunlop} M.,  {Baumgartel} K.,   {Tarasov} V.,
  1999, \mn@doi [\jgr] {10.1029/1998JA900143}, \href
  {http://adsabs.harvard.edu/abs/1999JGR...104.6763S} {104, 6763}

\bibitem[\protect\citeauthoryear{{Schaeffer}, {Winske}, {Larson}, {Cowee},
  {Constantin}, {Bondarenko}, {Clark}  \& {Niemann}}{{Schaeffer}
  et~al.}{2017}]{schaeffer17}
{Schaeffer} D.~B.,  {Winske} D.,  {Larson} D.~J.,  {Cowee} M.~M.,  {Constantin}
  C.~G.,  {Bondarenko} A.~S.,  {Clark} S.~E.,   {Niemann} C.,  2017, \mn@doi
  [Physics of Plasmas] {10.1063/1.4978882}, \href
  {http://adsabs.harvard.edu/abs/2017PhPl...24d1405S} {24, 041405}

\bibitem[\protect\citeauthoryear{{Sentman}, {Edmiston}  \& {Frank}}{{Sentman}
  et~al.}{1981}]{sentman81}
{Sentman} D.~D.,  {Edmiston} J.~P.,   {Frank} L.~A.,  1981, \mn@doi [\jgr]
  {10.1029/JA086iA09p07487}, \href
  {http://adsabs.harvard.edu/abs/1981JGR....86.7487S} {86, 7487}

\bibitem[\protect\citeauthoryear{{Terasawa}}{{Terasawa}}{1988}]{terasawa88}
{Terasawa} T.,  1988, \mn@doi [Computer Physics Communications]
  {10.1016/0010-4655(88)90226-3}, \href
  {http://adsabs.harvard.edu/abs/1988CoPhC..49..193T} {49, 193}

\bibitem[\protect\citeauthoryear{{Weibel}}{{Weibel}}{1959}]{weibel59}
{Weibel} E.~S.,  1959, \mn@doi [Physical Review Letters]
  {10.1103/PhysRevLett.2.83}, \href
  {http://adsabs.harvard.edu/abs/1959PhRvL...2...83W} {2, 83}

\bibitem[\protect\citeauthoryear{{Weidl}, {Jenko}, {Teaca}  \&
  {Schlickeiser}}{{Weidl} et~al.}{2015}]{weidl15}
{Weidl} M.~S.,  {Jenko} F.,  {Teaca} B.,   {Schlickeiser} R.,  2015, \mn@doi
  [\apj] {10.1088/0004-637X/811/1/8}, \href
  {http://adsabs.harvard.edu/abs/2015ApJ...811....8W} {811, 8}

\bibitem[\protect\citeauthoryear{{Weidl}, {Winske}, {Jenko}  \&
  {Niemann}}{{Weidl} et~al.}{2016}]{weidl16}
{Weidl} M.~S.,  {Winske} D.,  {Jenko} F.,   {Niemann} C.,  2016, \mn@doi
  [Physics of Plasmas] {10.1063/1.4971231}, \href
  {http://adsabs.harvard.edu/abs/2016PhPl...23l2102W} {23, 122102}

\bibitem[\protect\citeauthoryear{{Wiener}, {Pfrommer}  \& {Oh}}{{Wiener}
  et~al.}{2017}]{wiener17}
{Wiener} J.,  {Pfrommer} C.,   {Oh} S.~P.,  2017, \mn@doi [\mnras]
  {10.1093/mnras/stx127}, \href
  {http://adsabs.harvard.edu/abs/2017MNRAS.467..906W} {467, 906}

\bibitem[\protect\citeauthoryear{{Winske} \& {Leroy}}{{Winske} \&
  {Leroy}}{1984}]{winske84}
{Winske} D.,  {Leroy} M.~M.,  1984, \mn@doi [\jgr] {10.1029/JA089iA05p02673},
  \href {http://adsabs.harvard.edu/abs/1984JGR....89.2673W} {89, 2673}

\bibitem[\protect\citeauthoryear{{Winske}, {Thomas}, {Omidi}  \&
  {Quest}}{{Winske} et~al.}{1990}]{winske90}
{Winske} D.,  {Thomas} V.~A.,  {Omidi} N.,   {Quest} K.~B.,  1990, \mn@doi
  [\jgr] {10.1029/JA095iA11p18821}, \href
  {http://adsabs.harvard.edu/abs/1990JGR....9518821W} {95, 18821}

\bibitem[\protect\citeauthoryear{{Zirakashvili} \& {Ptuskin}}{{Zirakashvili} \&
  {Ptuskin}}{2008}]{zirakashvili08}
{Zirakashvili} V.~N.,  {Ptuskin} V.~S.,  2008, \mn@doi [\apj] {10.1086/529580},
  \href {http://adsabs.harvard.edu/abs/2008ApJ...678..939Z} {678, 939}

\bibitem[\protect\citeauthoryear{{Zweibel} \& {Everett}}{{Zweibel} \&
  {Everett}}{2010}]{zweibel10}
{Zweibel} E.~G.,  {Everett} J.~E.,  2010, \mn@doi [\apj]
  {10.1088/0004-637X/709/2/1412}, \href
  {http://adsabs.harvard.edu/abs/2010ApJ...709.1412Z} {709, 1412}

\makeatother
\end{thebibliography}

\end{document}